\documentclass[aps,pra,twocolumn]{revtex4-1}
\usepackage{
amsmath,
amssymb,
mathrsfs,
braket,
graphicx,
enumitem
}
\AtBeginDocument{\renewcommand{\natexlab}[1]{#1}}
\begin{document}
\begin{titlepage}
\title{Time Operators and Time Crystals: Self-Adjointness by Topology Change}
\author{
K. Nakatsugawa$^{1,2}$, 
T. Fujii$^{2,3}$, 
A. Saxena$^{4}$, 
and 
S. Tanda$^{1,2}$}
\vspace*{.2in}
\affiliation{
$^1$Department of Applied Physics, Hokkaido University, Sapporo 060-8628, Japan
\\
$^2$Center of Education and Research for Topological Science and Technology, Hokkaido University, Sapporo 060-8628, Japan
\\
$^3$Department of Physics, Asahikawa Medical University, Asahikawa 078-8510, Japan
\\
$^4$Center for Nonlinear Studies and Theoretical Division, Los Alamos National Laboratory, Los Alamos, New Mexico 87545, USA
}
\date{\today}
\begin{abstract}
We investigate time operators in the context of quantum time crystals in ring systems. A generalized commutation relation called the \emph{generalized weak Weyl relation} is used to derive a class of self-adjoint time operators for ring systems with a periodic time evolution: The conventional Aharonov-Bohm time operator is obtained by taking the infinite-radius limit. Then, we discuss the connection between time operators, time crystals and real-space topology. We also reveal the relationship between our time operators and a $\mathcal{PT}$-symmetric time operator. These time operators are then used to derive several energy-time uncertainty relations.
\end{abstract}
\nopagebreak
\maketitle
\end{titlepage}

%%%%%%%%%%%%%%%%%%%%%%%%%%
\section{Introduction}
%%%%%%%%%%%%%%%%%%%%%%%%%%
In the framework of standard quantum mechanics, time is not an observable but just a parameter:
%\textcolor{red}{For instance, one may seek to measure the time at which a quantum system is in a certain state, analogous to the measurement of position via the operator $\hat x$. However, the existence of such arrival time operators in quantum mechanics is an open problem.} 
One of the reasons is the difficulty to define a self-adjoint time operator $\hat T$ conjugate to a Hamiltonian operator $\hat H$ which satisfies the canonical commutation relation \cite{note1,ReedAndSimon,
Arrival_Time_in_Quantum_Mechanics}
\begin{equation}
[\hat H,\hat T]=i\hbar. \label{CCR1}
\end{equation}
Such time operators are necessary to derive time-energy uncertainty relations as well as for the unification of space and time in quantum mechanics. However, how to define time operators is still an open problem. 

\begin{figure}
\includegraphics[width=0.75\linewidth]{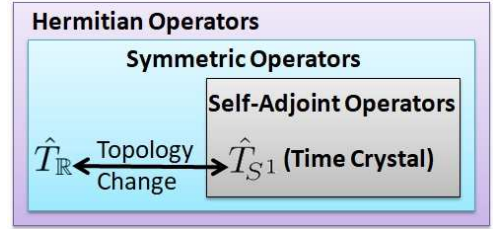}
\caption{Conventionally, physical observables have to be represented by self-adjoint operators. The Aharonov-Bohm time operator $\hat T_\mathbb{R}$ [Eq. \eqref{TAB}] is a symmetric operator which describes the arrival time of a free particle on a line $\mathbb R$ \cite{ABTimeOperator}. We derive the self-adjoint operator $\hat T_{S^1}$ [Eq. \eqref{TABonS1}] for a free particle on a ring system $S^1$. $\hat T_\mathbb{R}$ is obtained by taking the infinite radius limit of $\hat T_{S^1}$.}
\label{Inclusion_relation_wo_PT}
\end{figure}

The difficulty to define time operators lies in the difference between self-adjoint operators and symmetric operators even though these are both Hermitian operators (Fig. \ref{Inclusion_relation_wo_PT}). The existence of orthogonal eigenstates and real eigenvalues is ensured for self-adjoint operators but not ensured for symmetric operators. So, observables in standard quantum mechanics have to be represented by self-adjoint operators \cite{ReedAndSimon} but most of the time operators which satisfy Eq. \eqref{CCR1} are symmetric non-self-adjoint operators \cite{Arrival_Time_in_Quantum_Mechanics}. For example, the Aharonov-Bohm time operator \cite{ABTimeOperator}
\begin{equation}
\hat T_\mathbb{R}=-\frac{m}{2}(\hat x\hat p^{-1}+\hat p^{-1}\hat x)\label{TAB}
\end{equation}
which describes the arrival time of a free particle on a line ($\mathbb R$) is a symmetric operator without orthonormal eigenstates \cite{Arrival_Time_in_Quantum_Mechanics}. Such objection was first raised by W. Pauli's well-known theorem \cite{Pauli1958}. 
E. Galapon disproved Pauli's theorem by specifying Pauli's implicit assumptions, then he proved that self-adjoint time operators which satisfy Eq. \eqref{CCR1} can be defined in confined systems \cite{Galapon451,CQTA}. The mathematical structure of time operators for a Hamiltonian with discrete eigenvalues was studied by Galapon \cite{Galapon451,GalaponDiscrete}, Arai and Matsuzawa \cite{AraiMatsuzawa}, and Arai \cite{AraiDiscrete}.

In this article, we consider the above problem in the context of quantum time crystals (QTC) in ring systems \cite{QTC, Wigner, WilczekFFLO, CDWQTC, Nozieres}. A QTC is a quantum mechanical state which spontaneously breaks time translation symmetry \cite{QTC, Wigner, WilczekFFLO, CDWQTC, Nozieres, Sacha,FTC, PhaseStructure, FTCTheory, DTCYao, Prethermal, FTCChoi, FTCZhang}. 
The periodic oscillation of a QTC seems to promote time from a parameter to a physical observable. So, we propose that QTC are candidates for constructing time operators.

In section \ref{SATOS1} we use a generalized commutation relation called the \emph{generalized weak Weyl relation} (GWWR) \cite{GWWR} to derive a class of self-adjoint time operators for a free particle in a ring system. The GWWR is required for a consistent quantization of ring systems. These time operators show a periodic time evolution intrinsic to ring systems. The conventional Aharonov-Bohm time operator is obtained by taking the infinite-radius limit.
%We conclude that our time operator indeed describes the periodic time evolution of a QTC in a ring system. 
Then we discuss the connection between the GWWR, time operators, time crystals, and real-space topology of the system.

Moreover, operators obeying $\mathcal{PT}$-symmetry have real eigenvalues if their eigenstates are $\mathcal{PT}$-symmetric as well \cite{Bender1998, Bender2007, Bender2015}, but the possibility of $\mathcal{PT}$-symmetric time operators has not been considered in literature. So, in section \ref{PTTOS1} we reveal the relationship between the self-adjoint time operators in section \ref{SATOS1} and a $\mathcal{PT}$-symmetric time operator with orthogonal eigenstates and real eigenvalues. This $\mathcal{PT}$-symmetric operator reduces to a non-Hermitian time operator \cite{Recami10} in the infinite-radius limit.
%We conclude that our time operators indeed describe the periodic time evolution of a QTC in a ring system.
Then, in section \ref{TEUncertainty}, we derive several energy-time uncertainty relations for self-adjoint and $\mathcal{PT}$-symmetric time operators. 
\section{Self-Adjoint Time Operators in $S^1$}
\label{SATOS1}
%%%%%%%%%%%%%%%%%%%%%%%%%%
\subsection{General Weak Weyl Relation and Real-Space Topology}
%%%%%%%%%%%%%%%%%%%%%%%%%%
There is a hierarchy of operators which satisfy stronger and weaker versions of the canonical commutation relation \cite{ReedAndSimon, WWE,Miyamoto,GWWR,AraiUWTO, Ohnuki,CARRUTHERSandNIETO, TanimuraS1}. Here, we consider the \emph{generalized weak Weyl relation} (GWWR) introduced by A. Arai \cite{GWWR}. As we exemplify below, the GWWR is necessary for a consistent quantization of ring systems. Let $\hat{A}$ be a symmetric operator on a Hilbert space $\mathcal{H}$, let $\hat B$ be a self-adjoint operator on $\mathcal{H}$ and $\hat{K}(s)$ ($s\in\mathbb{R}$, $s$ may denote time, position, or some other variable) be a bounded self-adjoint operator on $\mathcal{H}$ with $D(\hat K(s))=\mathcal{H}$ ($D(\cdot)$ denotes operator domain), $\forall s\in\mathbb R$. We say that $(\hat A,\hat B,\hat K)$ obeys the GWWR in $\mathcal{H}$ if for all $\ket{\psi}\in D(\hat A)$ and for all $s\in\mathbb R$ we have $e^{-is\hat B/\hbar}\ket{\psi}\in D(\hat A)$ and
\begin{equation}
\hat A e^{-is\hat B/\hbar}\ket{\psi} = e^{-is\hat B/\hbar}(\hat A + \hat K(s))\ket{\psi}.
\label{GWWRDef}
\end{equation}
$\hat K(s)$ is called the commutation factor of the GWWR. 
%Eq. \eqref{GWWRDef} reduces to the \emph{weak Weyl relation} for $\hat K(t)=\pm t.$ \cite{WWE,Miyamoto}.
% and Eq. \eqref{GWCCR} reduces to the canonical commutation relation$[\hat A,\hat B]\ket{\psi}=\pm i\hbar\ket{\psi}$.
$\hat A$ and $\hat B$ may represent position, momentum, angular momentum, Hamiltonian, time operator, etc.
Moreover, if $\hat K(s)$ is differentiable with respect to $s$, then we can differentiate both sides of Eq. \eqref{GWWRDef} and set $s=0$ to obtain the \emph{generalized canonical commutation relation} (GCCR) for all $\ket{\psi}\in D(\hat A\hat B)\cap D(\hat B\hat A)$
\begin{align}
[\hat A,\hat B]\ket{\psi}=i\hbar\hat K'(0)\ket{\psi}, \label{GWCCR}
\end{align}
where $'$ denotes derivative with respect to $s$.  For $\hat K(s)=\pm s$ Eq. \eqref{GWWRDef} reduces to the \emph{weak Weyl relation} \cite{WWE,Miyamoto} and Eq. \eqref{GWCCR} reduces to the canonical commutation relation$[\hat A,\hat B]\ket{\psi}=\pm i\hbar\ket{\psi}$.
We conjecture that the commutation factor $\hat K(s)$, in general, depends on the real-space topology of a quantum system. Several examples are given below.
\subsubsection{One-dimensional system}
The Hilbert space of a one-dimensional system is $\mathcal H=L^2(\mathbb R)$. The position operator $\hat x$ and the momentum operator $\hat p$ are both self-adjoint. For $(\hat A,\hat B,\hat K(s))=(\hat x,\hat p, s)$ we have $[\hat x,\hat p]=i\hbar$. For $(\hat A,\hat B,\hat K(t))=(\hat T_\mathbb R,\hat H, -t)$ we have $[\hat H,\hat T_\mathbb R]=i\hbar$, where $\hat H=\hat p^2/2m$ and $\hat T_\mathbb R$ is the Aharonov-Bohm time operator defined in Eq. \eqref{TAB}. 
\subsubsection{Confined system}\label{confinedsystem}
Galapon \cite{Galapon451} and Galapon, Caballar, and Bahague \cite{CQTA} considered time operators in confined systems with the Hilbert space $\mathcal H=L^2([-l,l])$ and with the boundary condition $\phi(-l)=e^{-2i\gamma}\phi(l)$, $\phi\in\mathcal H$, $|\gamma|<\pi$. Let us consider a free particle with the Hamiltonian $\hat H_\gamma=\hat p_\gamma^2/2m$.
%$\hat x$ and $\hat p_\gamma$ are self-adjoint for all $\gamma$. 
If $\gamma\ne0$ then $p=0$ is \emph{not} an eigenvalue of $\hat p_\gamma$, so $\hat p_\gamma^{-1}$ is a bounded self-adjoint operator. If $\gamma=0$ (periodic boundary condition) then $p=0$ is an eigenvalue of $\hat p_\gamma$, so $\hat p_\gamma^{-1}$ is well-defined only if the zero-momentum state $\ket{p=0}$ is removed using projection operators \cite{Galapon451,CQTA}. In both cases, it is possible to define a self-adjoint time operator $\hat T_\gamma=-\frac{m}{2}(\hat x\hat p_\gamma^{-1}+\hat p_\gamma^{-1}\hat x)$ which satisfies the canonical commutation relation $[\hat H_\gamma,\hat T_\gamma]=i\hbar$. This commutation relation is well-defined within a restricted domain. 
\\
\indent Our main concern is as follows. Position operators are defined as a set of operators whose eigenvalues have one-to-one correspondence with points on a manifold. But in the case $\gamma=0$ (periodic boundary condition) $\hat x$ is a multivalued operator, hence it is not  suitable as a position operator \cite{Ohnuki, CARRUTHERSandNIETO, TanimuraS1}. For instance, any continuous wave function $\psi$ of a ring system must satisfy the periodic boundary condition but $\hat x\psi$ is not periodic. Moreover, if $\hat x$ is not used with care, then many contradictions occur. For example, the expectation value of the canonical commutation relation with momentum eigenstates $\psi_p$ (including the ground state $\psi_0$) leads to two equivalent equations
$\braket{\psi_p|[\hat x,\hat p_0]|\psi_p}=p(\braket{\psi_p|\hat x|\psi_p}-\braket{\psi_p|\hat x|\psi_p})=0$ and $\braket{\psi_p|[\hat x,\hat p_0]|\psi_p}=\braket{\psi_p|i\hbar|\psi_p}=i\hbar$.
These results lead to the contradiction $0=i\hbar$. Similarly, one can verify that $\braket{\psi_p|[\hat H_0,\hat T_0]|\psi_p}=0$ for $p\ne0$. These facts suggest that the pairs $(\hat x,\hat p_0)$ and  $(\hat T_0,\hat H_0)$ are very sensitive to the domain on which their commutation relation is applied. In other words, the canonical commutation relation imposes a restriction on possible wave functions of a ring system \cite{Galapon451, AraiMatsuzawa}. 
So, how can we obtain position operators, time operators and commutation relations for general periodic functions?
This problem can be solved with periodic position operators as we show below.
\subsubsection{Ring systems}
Let $\mathcal H=L^2([-\pi,\pi])$ be the Hilbert space of ring systems with the periodic boundary condition $\psi(\theta)=\psi(\theta+2\pi)$, where $\theta=x/R$ is the angular coordinate.
As we exemplified above, how to quantize ring systems has been an important problem in the formulation of quantum mechanics. Previous studies to solve this problem include refs. \cite{Ohnuki, CARRUTHERSandNIETO, TanimuraS1}. In all cases, the solution was to use periodic angle variables such as $f(\theta)=\theta\ \text{mod}\ 2\pi$, $f(\theta)=\cos\theta$, $f(\theta)=\sin\theta$, and $f(\theta)=e^{i\theta}$. In fact, these different formulations have the same mathematical structure, namely the GWWR. 
Instead of the multivalued position operator $\hat x$ or the angle operator $\hat \theta=\hat x/R$ we use a periodic position operator $\hat f$ such that $\hat f\ket{\theta}=f(\theta)\ket{\theta}$ with $f(\theta+2\pi)=f(\theta)$. We also define $\hat f_s=e^{i\hat\pi_\theta s/\hbar}\hat f e^{-i\hat\pi_\theta s/\hbar}$ with $\hat f_s\ket{\theta}=f(\theta+s)\ket{\theta}$, $\hat\pi_\theta=R\hat p$ is the canonical angular momentum operator. Then, from the GWWR with $\hat A=\hat f$ and $\hat B=\hat\pi_\theta$ we obtain $\hat K(s)=\hat f_s-\hat f$, $\hat K'(s)=\hat f'_s=d\hat f_s/ds$, and these operators satisfy the GCCR
\begin{equation}
[\hat \pi_\theta,\hat f]=-i\hbar\hat f'_s\big|_{s=0}.\label{GCCRPeriodic}
\end{equation} 
Any physical periodic operators can be constructed using Fourier series
\begin{align}
\hat f\ket{\theta}&=f(\theta)\ket{\theta}
=\sum_{n=-\infty}^\infty c_ne^{in\theta}\ket{\theta}
=\sum_{n=-\infty}^\infty c_n\hat W^n\ket{\theta}.\nonumber
%\\
%&\Rightarrow \hat f=\sum_{n=-\infty}^\infty c_n\hat W^n
\end{align}
Here, we have defined the unitary position operator $\hat W=\widehat{e^{i\theta}}$, $\hat W^\dagger=\hat W^{-1}$ \cite{Ohnuki,TanimuraS1}. $\hat\pi_\theta$ and $\hat W$ satisfy the following eigenvalue equations and commutation relation in $\mathcal H$
\begin{align}
\hat W^n\ket{\theta}&=e^{in\theta}\ket{\theta},
\\
\hat\pi_\theta\ket{l}&=l\hbar\ket{l},\label{PiEigeneq}
\\
\hat W^{n}\ket{l}&=\ket{l+n}\label{WRaising}
\\
[\hat\pi_\theta,\hat W^n]&=n\hbar\hat W^n\label{WAlgebragen}.
\end{align}
Note that $l$ and $n$ are integers and one can verify that $\hat W^{-n}=\hat W^{\dagger n}$.
%$\hat C$ and $\hat S$ satisfy similar commutation relations. 
From this definition with Eq. \eqref{GCCRPeriodic} and Eq. \eqref{WAlgebragen} we obtain
\begin{align*}
\hat f'_s\big|_{s=0}
%&=\sum_{n=-\infty}^\infty c_n\hat W^{n\prime}_s
=\sum_{n=-\infty}^\infty c_n\frac{i}{\hbar}[\hat\pi_\theta,\hat W^n]
=\sum_{n=-\infty}^\infty c_nin\hat W^n.
%=\frac{i}{\hbar}[\hat\pi_\theta,f_s]
\end{align*}
%which reproduces Eq. \eqref{GCCRPeriodic} as $s\to0$. 
Besides, it is a known fact from spectral analysis that the real and imaginary parts of a bounded operator (i.e. an operator whose spectrum is bounded from above and below) are self-adjoint operators. $\hat W^n$s are bounded operators, so $\hat f$ is a bounded self-adjoint operator if $c_n=c_{-n}^\ast$. 
As specific examples of $\hat f$ we may take the self-adjoint sine operator $\hat S$ and the self-adjoint cosine operator $\hat C$ to specify a point on a ring \cite{CARRUTHERSandNIETO, TanimuraS1}.
\begin{align}
\hat C&=\frac{\hat W+\hat W^\dagger}{2}\quad (c_n=c_{-n}^\ast=\frac{1}{2}\delta_{n,1}),\label{Cdef}
\\
\hat S&=\frac{\hat W-\hat W^\dagger}{2i}\quad (c_n=c_{-n}^\ast=\frac{i}{2}\delta_{n,1}),\label{Sdef}
\end{align}
Another important operator is the single-valued periodic angle operator $\hat\Theta\ket{\theta}=(\theta \mbox{ mod}2\pi)\ket{\theta}$ \cite{CARRUTHERSandNIETO}
\begin{align}
\hat\Theta&=\sum_{n=1}^\infty\frac{(-1)^{n+1}}{in}(\hat W^n-\hat W^{-n})=-i\text{Log}(\hat W)
\label{Thetadef}
\end{align}
which is the operator version of the ``sawtooth function" $\Theta(\theta)=2\sum_{n=1}^\infty(-1)^{n+1}n^{-1}\sin(n\theta)=-i\text{Log}(e^{i\theta})$, i.e. the angle $\theta$ restricted in the region $[-\pi,\pi]$.
These operators satisfy the commutation relations
\begin{align}
%\hat W&=\hat C+i\hat S.
%\\
[\hat\pi_\theta,\hat C]&=i\hbar\hat S,\label{CSAlgebra}
\\
[\hat\pi_\theta,\hat S]&=-i\hbar\hat C,\label{SCAlgebra}
\\
[\hat\Theta,\hat\pi_\theta]&=i\hbar\{1-\delta(\hat\Theta+\pi)\}\label{ThetaAlgebra}
%[\hat C,\hat S]&=0,
%\\
\end{align}
\indent Now, one of the requirements to define time operators is that they satisfy a commutation relation similar to that of position and momentum: This requirement is necessary to derive time-energy uncertainty relations as well as for the unification of space and time in quantum mechanics. In this context, we propose that the quantization of other physical observables in ring systems, including time intervals and events, should be done based on the GWWR. One can either use $[\hat H,\hat T]=i\hbar$ with a restricted domain or consider a generalized time operator with a larger domain (i.e. many other physical states) such that $[\hat H,\hat T]=i\hbar \hat K'(0)$. There is no obvious reason that a time operator must satisfy the former case, especially when boundary conditions play an important role. If periodic position operators $\hat f$ are used instead of $\hat\theta$, then time operators are expected to obey a commutation relation similar to Eq. \eqref{GCCRPeriodic}. 
%%%%%%%%%%%%%%%%%%%%%%%%%%
\subsection{Derivation of Generalized Self-Adjoint Time Operators}
%%%%%%%%%%%%%%%%%%%%%%%%%%
Using the quantization of ring systems based on the GWWR and the GCCR, we show that there exists a class of generalized self-adjoint time operators in ring systems with $\hat K'(0)=\hat f'_s\big|_{s=0}$. We take Eq. \eqref{GCCRPeriodic} as a starting point to derive these operators. Consider a free particle with moment of inertia $I$ and Hamiltonian $\hat H=\frac{\hat\pi_\theta^2}{2I}$. 
Consider the following commutation relation
\begin{align*}
%\frac{d}{dt}\hat W^n(t)=\frac{i}{\hbar}[\hat H,\hat W^n(t)]=i\hat W^n(t)\frac{n}{2I}(2\hat\pi_\theta+n\hbar).
[\hat H,\hat W^n]\ket{\psi}
&=[\hat H,\hat W^n]\sum_{l=-\infty}^\infty\ket{l}\braket{l|\psi}
\\
&=\sum_{l=-\infty}^\infty(E_{l+n}-E_l)\ket{l+n}\braket{l|\psi},
%=\hat W^n\frac{n\hbar}{2I}(2\hat\pi_\theta+n\hbar).
\end{align*}
where we used Eq. \eqref{WRaising}, $\hat H\ket{l}=E_l\ket{l}$ and the identity operator $\mathbf 1=\sum_{l=-\infty}^\infty\ket{l}\bra{l}$.
The desired time operator is obtained if we can get rid of $E_{l+n}-E_l=\frac{(2l+n)n\hbar^2}{2I}$. If $n$ is an odd integer then $l=-\frac{n}{2}$ is not an eigenvalue of $\hat\pi_\theta$, so $1/(2l+n)$ is bounded. On the other hand, if $n$ is an even integer, we can use the projection operator $\mathscr P_{-n/2}$ to remove the state $\ket{-\frac{n}{2}}$. 
In both cases we can define the time operator $\hat T$ as
\begin{align*}
\hat T\ket{\psi}&=-\sum_{n,l=-\infty}^\infty c_{n}n\hbar\frac{(1-\delta_{l,-\frac{n}{2}})}{E_{l+n}-E_l}\ket{l+n}\braket{l|\psi},
\\
\ket\psi\in D(\hat T)&=\bigg\{\sum_{l=-\infty}^\infty a_{nl}\ket{l}:\sum_{l=-\infty}^\infty|a_{nl}|^2<\infty\bigg\}.
\end{align*}
This kind of time operator has been studied in refs. \cite{GalaponDiscrete, AraiMatsuzawa, AraiDiscrete}.
The commutation relation between this time operator and the Hamiltonian gives
\begin{align*}
[\hat H,\hat T]\ket\psi
&=-\sum_{n,l=-\infty}^\infty c_{n}n\hbar(1-\delta_{l,-\frac{n}{2}})\ket{l+n}\braket{l|\psi}
\\
&=-\sum_{n=-\infty}^\infty c_{n}n\hbar \hat W^n\ket{\psi}+\sum_{n=-\infty}^\infty c_{2n}2n\hbar\ket{n}\braket{-n|\psi}
\\
&=i\hbar\hat f'_s\big|_{s=0}\ket{\psi}+\sum_{n=-\infty}^\infty c_{2n}2n\hbar\ket{n}\braket{-n|\psi},
\\
\ket{\psi}&\in D(\hat H\hat T)\cap D(\hat T\hat H).
\end{align*}
Here, ${\ket{n}}$ are linearly independent, hence the last term with $c_{2n}$ vanishes only if $c_{2n}\braket{-n|\psi}=0$ for all $n$. %If $c_{2n}\ne0$ for all $n$, then $\braket{-n|\psi}=0$ implies the trivial solution $\ket{\psi}=0, i.e. D(\hat H\hat T)\cap D(\hat T\hat H)=\{0\}$. If $c_{2n}\ne0$ for finite number of $n$, then a sufficient condition is $\ket{\psi}=\prod_{\{n=\text{even}:c_{n}\ne0\}}\mathscr P_{-n/2}\sum_{l=-\infty}^\infty a_l\ket{l}$ where $\mathscr P_{l}=(1-\ket{l}\bra{l})$, $\mathscr P_{l}^2=\mathscr P_{l}$.
Finally, we can define the bounded self-adjoint operator which has the dimension of time
\begin{align*}
\hat\mu_n=\left\{
\begin{matrix}
\frac{2I}{2\hat\pi_\theta+n\hbar}&,\ n=\mbox{odd}
\\
\mathscr P_{-n/2}\frac{2I}{2\hat\pi_\theta+n\hbar}\mathscr P_{-n/2}&,\ n=\mbox{even}
\end{matrix}
\right.
\end{align*}
and write our time operator as
\begin{align}
\begin{split}
\hat T&=-\sum_{n}c_n\hat W^n\hat\mu_n
\\
&=-\sum_{n}\frac{(c_n\hat W^n\hat\mu_n+c_n^\ast\hat\mu_n\hat W^{-n})}{2},
\end{split}
\label{TODef}
\\
&[\hat H,\hat T]\ket{\psi}=i\hbar\hat f'_s\big|_{s=0}\ket{\psi},\label{CCRGen}
\\
&\ket{\psi}\in\bigg\{\ket{\psi}\in D(\hat T):c_{2n}\braket{-n|\psi}=0, \forall n\bigg\}.\nonumber
\end{align}
The symmetrized form in the second line is obtained from the fact that $\hat W^n\hat\mu_n=\hat\mu_{-n}\hat W^n$ and $c_n^\ast=c_{-n}$. This time operator has the following properties:
\\
1. $\hat T$ is a bounded symmetric operator, hence it is self-adjoint. Another reason that this operator is self-adjoint is that the Hamiltonian has discrete energy spectrum \cite{Galapon451, GalaponDiscrete, AraiMatsuzawa, AraiDiscrete}. Discreteness of the energy spectrum is very important but, as we have seen, it is not the only factor that determines the time operators.
\\
2. $\hat T$ satisfies the GWWR \eqref{GWWRDef} with $(\hat A,\hat B,\hat K(t))=(\hat T,\hat H,\hat T(t)-\hat T)$, $\hat T(t)=e^{i\hat Ht/\hbar}\hat T e^{-i\hat Ht/\hbar}$. One can readily show that $d\hat K(t)/dt|_{t=0}=-\hat f'_s\big|_{s=0}$ and obtain the correct generalized commutation relation.
\\
3. We note that $\hat T$ is not unique because, for any self-adjoint operator $\hat F$ which commutes with the Hamiltonian, $\hat T$ and the new operator $\hat T+\hat F$ satisfy the same commutation relation with $\hat H$. Such operators can be added if necessary. However, if $\hat f$ is prescribed then $\hat T$ is unique up to addition of $\hat F$. Which $\hat f$ we choose depends on physical basis, i.e. on the phenomena (such as time of arrival, time crystal or other events) that we want to observe.
\\
4. If we choose $\hat f$ such that $R\hat f\to \hat x$ in the infinite-volume limit $R\to\infty$, then the commutation relation Eq. \eqref{CCRGen} reduces to the canonical commutation relation \eqref{CCR1}. This limit must be taken for $\ket{\psi}$ in the domain $\mathcal D\subset D(\hat H\hat T)\cap D(\hat T\hat H)$ such that $\ket{\psi}\in\mathcal D$ is also square-integrable for $R\to\infty$. This domain includes the Gaussian-like minimum-uncertainty state considered by S. Tanimura \cite{TanimuraS1}.
\\
5. $\hat T$ is defined as a linear combination of non-Hermitian operators $\hat W^n\mu_n$. These operators have space-time reversal ($\mathcal{PT}$) symmetry. The significance of $\mathcal{PT}$-symmetric time operators is discussed in section \ref{PTTOS1}.
\\
\indent Now, let us consider some special cases and discuss their physical significance. If $c_n=c_{-n}^\ast=\frac{i}{2}\delta_{n,1}$ we obtain
\begin{align}
\hat T_{S^1}&=\frac{1}{2i}[\hat\mu\hat W^\dagger-\hat W\hat\mu]=\text{Im}[\hat\mu\hat W^\dagger],
\label{TABonS1}
\end{align}
which satisfies the commutation relation
\begin{align}
[\hat H,&\hat T_{S^1}]=i \hbar\hat C.
\label{TImAlgebra}
\end{align}
%\subsection{Infinite Radius Limit of $\hat T_{S^1}$}

Note that $\hat T_{S^1}$ reduces to $\hat T_\mathbb{R}$ in the infinite-radius limit $R\to\infty$. Using the identity operators $\int_{-\pi}^\pi d\theta\ket{\theta}\bra{\theta}=\mathbf 1$, $\int_{-\infty}^\infty dx\ket{x}\bra{x}=\mathbf 1$, $\sum_{l\in\mathbb Z}\ket{l}\bra{l}=\mathbf 1$ and $\int_{-\infty}^\infty dk\ket{k}\bra{k}=\mathbf 1$ with $x=R\theta$, $k=l/R$, $p=\hbar k$, $\braket{\theta|l}=\braket{x|k}$, $I=mR^2$, $\mu_{l}=\braket{l|\hat\mu_1|l}$, $e^{i\theta}\approx 1+i\theta$ and $\mu_l/R\to m/p$ as $R\to\infty$ we obtain
\begin{align*}
\hat T_{S^1}&\approx-\text{Im}\left[\sum_{l\in\mathbb Z}\int_{-\pi}^\pi d\theta\ket{\theta}\bra{\theta}(1+i\theta)\mu_{l}\ket{l}\bra{l}\right]
\\
&=-\text{Re}\left[\sum_{l\in\mathbb Z}\int_{-\pi}^\pi d\theta\ket{\theta}\bra{\theta}\theta\mu_{l}\ket{l}\bra{l}\right]
\\
&\to -\text{Re}\left[\int_{-\infty}^{\infty}dk\int_{-\infty}^{\infty}dx\ket{x}\bra{x}\frac{mx}{\hbar k}\ket{k}\bra{k}\right]
\\
&=-\frac{m}{2}(\hat x\hat p^{-1}+\hat p^{-1}\hat x)=\hat T_\mathbb R.
\end{align*}
This limit must be taken with wave functions in some appropriate domain (as explained in Property 4. above). 
A sufficient condition for this limit is that the wave functions are well localized around $\theta=0$ so that the contribution from $\theta\sim R$ can be ignored.
Similarly, one can show that $R\hat{\mathcal S}\to \hat x$, $\hat{\mathcal C}\to 1$ as $R\to\infty$, hence
\begin{align*}
&\hat T_{S^1}\to \hat T_\mathbb{R},
\\
&[\hat S,\hat\pi_\theta]=i\hbar \hat C\to[\hat x,\hat p]=i\hbar,
\\
&[\hat H,\hat T_{S^1}]=i\hbar\hat C\to[\hat H,\hat T_\mathbb{R}]=i\hbar,
\end{align*}
as $R\to\infty$. Therefore, we conclude that $\hat T_{S^1}$ is a self-adjoint analogue of the Aharonov-Bohm time operator $\hat T_\mathbb{R}$ in $S^1$.
\\
If $c_n=c_{-n}^\ast=\frac{1}{2}\delta_{n,1}$ we can define the time operator $\hat T_{S^1}^\text{Re}$
\begin{equation}
\hat T_{S^1}^\text{Re}=\hat\mu_1-\frac{1}{2}[\hat W\hat\mu_1+\hat\mu_1\hat W^\dagger]\label{TREonS1}
\end{equation}
which satisfies the commutation relation
\begin{equation}
[\hat H,\hat T_{S^1}^\text{Re}]=-i\hbar\hat S.\label{TReAlgebra}
\end{equation}
Note that $\hat\mu_1$ in Eq. \eqref{TREonS1}, which commutes with the Hamiltonian $\hat H$, was included so that the matrix elements of $\hat T_{S^1}^\text{Re}$ do not diverge in the limit $R\to\infty$. The physical significance of $\hat T_{S^1}^\text{Re}$ can be understood by taking the large radius limit. Using $I=mR^2$, $l=Rk=Rp/\hbar$ and $\theta=x/R$, we find that $\hat T_{S^1}^\text{Re}$ has matrix elements
\begin{align*}
\braket{\theta|\hat T_{S^1}^\text{Re}|l}&=\frac{2I}{\hbar}\frac{(1-2l+2l \cos\theta-i \sin\theta)}{1-4l^2}\braket{\theta|l}
\\
&=-\frac{1}{4\pi}\frac{\lambda_\text{dB}}{v}\braket{\theta|l}+O(R^{-1}).
\end{align*}
In the second line we did a Taylor expansion for $\theta=x/R\ll 1$. Here $\lambda_\text{dB}=2\pi\hbar/p$ is the de Broglie wavelength and $v=p/m$ is the group velocity of the particle. The term $\lambda_\text{dB}/v$ describes a ``matter wave clock", i.e. because of the periodicity of de Broglie wavelength, a particle moving with a fixed velocity $v$ has an internal clock with period $\lambda_\text{dB}/v$ \cite{deBroglie,LanScience}. 
\\
Note that $\hat S$ in Eq. \eqref{TReAlgebra} vanishes as $R\to\infty$. So, although $T_{S^1}^\text{Re}$ is a time operator which satisfies the GWWR, it is not a time operator in the sense of Eq. \eqref{CCR1}.
\\
The third example is for the periodic angle operator $\hat \Theta$ defined in Eq. \eqref{Thetadef}. The corresponding time operator is defined as
\begin{align}
\hat T_\Theta=-\sum_{n=1}^\infty\frac{(-1)^{n+1}}{in}(\hat W^n\hat\mu_n-\hat\mu_n\hat W^{-n})
\end{align}
which satisfies the commutation relation
\begin{align}
[\hat H,\hat T_\Theta]=-i\hbar\{1-\delta(\hat\Theta+\pi)\}.\label{TThetaAlgebra}
\end{align}
The delta function $\delta(\hat\Theta+\pi)$ has a contribution only if $\theta+(2n+1)\pi=0$. Otherwise $\hat T_\Theta$ satisfies the canonical commutation relation and it is equivalent to Galapon's time operator $\hat T_0$ for periodic system. Therefore, $\hat T_\Theta$ is interpreted as a time-of-arrival operator.

%%%%%%%%%%%%%%%%%%%%%%%%%%
\subsection{Connection to Time Crystals}
\label{TC}
%%%%%%%%%%%%%%%%%%%%%%%%%%
How are these time operators connected to QTC? In our previous model of QTC, the Heisenberg operator $\hat C(t)=e^{i\hat Ht/\hbar}\hat Ce^{-i\hat Ht/\hbar}$ describes the local oscillation of an incommensurate charge density wave: this oscillation is intrinsic to a ring system and it is used to model a QTC \cite{CDWQTC}.
%This operator is obtained as in Eq. \eqref{TImAlgebra}. 
Other periodic operators $\hat f$ can also be used to model QTC in ring systems: Note that $\hat W^{n}$, as a momentum raising operator, can be written as $\hat W^{n}=\sum_{l}\ket{l+n}\bra{l}$. Then, because $(E_{l+n}-E_l)/(E_{0+n}-E_0)=2l+n$ is an integer, the Heisenberg operators $\hat W(t)=e^{i\hat Ht/\hbar}\hat We^{-i\hat Ht/\hbar}$, $\hat f(t)=e^{i\hat Ht/\hbar}\hat fe^{-i\hat Ht/\hbar}$ and $\hat T(t)=e^{i\hat H t/\hbar}\hat Te^{-i\hat H t/\hbar}$ have a periodic time evolution with period $P=2\pi(E_{1}-E_0)^{-1}=4\pi I/\hbar$:
\begin{align*}
\hat W&=\sum_{l=-\infty}^\infty\ket{l+n}\bra{l}e^{it(E_{l+n}-E_l)/\hbar}
\\
&=\sum_{l=-\infty}^\infty\ket{l+n}\bra{l}e^{i(t+P)(E_{l+n}-E_l)/\hbar}
\\
&=\hat W(t+P),
\\
\\
\Rightarrow\hat T(t)&=\hat T(t+P)\mbox{ and }\hat f(t)=\hat f(t+P).
\end{align*}
The period $P$ diverges as $R\to\infty$, so this periodicity is intrinsic to ring systems. In fact, for a one-dimensional free particle, it is clear from $\hat x(t)=\hat x+\hat pt/m$ that
\begin{align}
\hat T_\mathbb{R}(t)=\hat T_\mathbb{R}+t
\end{align}
is not periodic.
The commutation factor $\hat K(t)=\hat T(t)-\hat T$ of a ring system is also periodic with a radius-dependent period. On the other hand, for a one-dimensional system ($\mathbb R$) we have $\hat K(t)=t$ which implies Eq. \eqref{CCR1}, hence time for a one-dimensional system is not necessarily periodic. Therefore, $\hat K(t)$ may be interpreted as a function which gives the ``temporal structure" of a quantum system. 

Because we are using a general mathematical structure (Eq. \eqref{GWWRDef}) to construct time operators, our work is not limited to ring systems but applies to other QTC models as well. We surmise $\hat K(t)$ characterizes the time-periodic evolution of a QTC. For our previous QTC model, this quantity is the charge density operator $\hat C(t)$. QTC states can also be realized in excited Floquet states \cite{Sacha,FTC, PhaseStructure,FTCTheory, DTCYao, Prethermal, FTCChoi, FTCZhang}. If a Floquet system were driven with a period  $P$, a Floquet time crystal (FTC) would return to its initial state after period $nP$ ($n$ is an integer), hence time translation symmetry is spontaneously broken.
So, we conjecture the following set of operators for a FTC
\begin{equation}
\begin{matrix}
[\hat H(t),\hat T(t)]&=\dot{\hat K}(t),
\\
\hat H(t+P)&=\hat H(t),
\\
\hat T(t+nP)&=\hat T(t),
\\
\hat K(t+nP)&=\hat K(t).
\end{matrix}
\end{equation}
For instance, the Hamiltonians  $\hat H(t)$ of most of the FTC models proposed so far are composed of Pauli matrices $\sigma_i$. If we set $\hat H(t+P)=\hat H(t)=\left\{\begin{matrix}\sigma _1,&0\leq t<\pi,\\0, &\pi\leq t<P\end{matrix}\right.$ as a prototype of a Floquet Hamiltonian then we can identify $\hat T(t)=U^\dagger(t)\sigma_2 U(t), \hat K(t)=\hat T(t)-T(0)$, and $\dot{\hat K}(t)=2U^\dagger (t) \sigma_3 U(t)$ with the unitary time evolution $U(t)=e^{-\frac{i}{\hbar} \int_0^t\hat H(t')dt'}$. In this case, $\hat T$ and $\hat K(t)$ are periodic with period $2P$. As a GCCR we obtain the $SU(2)$ commutation relation $[\sigma_1,\sigma_2]=2i\sigma_3$. Further investigation of this conjecture is left for future study. 
\\
Therefore, it seems like time operators and time crystals are interrelated. The periods of a QTC seems to promote time from a parameter to a physical quantity. So, QTC are promising systems to define time operators. 
In spite of recently proposed models of QTC in excited states \cite{Sacha,FTC, PhaseStructure,FTCTheory, DTCYao, Prethermal, FTCChoi, FTCZhang}, the original idea of a QTC is to define a dynamical \emph{ground state} which breaks time translation symmetry \cite{QTC}. How to construct a QTC ground state is a very important open problem in quantum mechanics, quantum field theory, condensed matter physics, and related fields, because it concerns the question of what a ground state is. The existence of QTC ground states have been criticized by showing that spontaneous braking of time translation symmetry at ground state does not occur in the infinite-volume limit \cite{Nozieres,Watanabe}. However, if the periodic boundary condition is not negligible, our results suggest how a time crystal can be defined in ring systems. We will have a QTC ground state if $\hat K(t)=\hat T(t)-\hat T$ and $d\hat K(t)/dt$ have periodic expectation value at ground state. Instead of a ring system we can also consider other systems with non-trivial real-space topologies. Once we have the appropriate mathematical structure to define a time operator (such as Eq. \eqref{GWWRDef}), then we can apply it to models of time crystals, possibly including ground states.

%%%%%%%%%%%%%%%%%%%%%%%%%%
\section{Extension to $\mathcal{PT}$-Symmetric Time Operators}
%%%%%%%%%%%%%%%%%%%%%%%%%%
\label{PTTOS1}
In the previous sections we focused on self-adjoint time operators because real eigenvalues and orthogonal eigenstates are not ensured for non-self-adjoint symmetric operators. Here, we extend time operators to $\mathcal{PT}$-symmetric operators for the following reasons. 
First, if $\mathcal{PT}$ symmetry is not spontaneously broken, then $\mathcal{PT}$-symmetric operators have real eigenvalues and bi-orthogonal eigenstates \cite{Bender1998, Bender2007, Bender2015, BOQM}. In other words, time operators do not have to be self-adjoint to have real-eigenvalues and orthogonal eigenstates. 
Second, we cannot completely distinguish from the canonical commutation relation and from the time-energy uncertainty relation if a time operator is self-adjoint or $\mathcal{PT}$-symmetric. This argument is elaborated in section \ref{TEUncertainty}. Therefore, it is possible that many time-operators are not self-adjoint (not even Hermitian) but $\mathcal{PT}$-symmetric.
The third reason is of academic interest: So far, the main emphasis of $\mathcal{PT}$-symmetric quantum mechanics is on the spectral properties of $\mathcal{PT}$-symmetric Hamiltonians. So, $\mathcal{PT}$-symmetric time operators open a door to study other non-Hermitian operators, which can give us real eigenvalues and (bi-)orthonormal eigenstates.
Therefore, in this section we define $\mathcal{PT}$ symmetry in ring systems. Then we derive a $\mathcal{PT}$-symmetric time operator. For simplicity we will focus only on the $\mathcal{PT}$-symmetric extension of \eqref{TABonS1}. Other $\mathcal{PT}$-symmetric operators can be defined likewise.

Before we define $\mathcal{PT}$-symmetric time operators it may be appropriate to discuss their validity as observables. Observables in quantum mechanics, by definition, are operators which can be measured experimentally. Self-adjoint operators are observables because real eigenvalues and orthogonal eigenstates are ensured. But, this does not mean that observables must be self-adjoint. In fact, $\mathcal{PT}$-symmetric operators can be measured \cite{Bender2015}. In addition, nowadays non-Hermitian operators can also be measured by methods like weak measurement. Therefore, in our opinion, $\mathcal{PT}$-symmetric operators are genuine observables which can be measured experimentally.
\subsection{$\mathcal{PT}$-Symmetry for Ring Systems}
For quantum mechanics in one-dimension $\mathbb R$, the parity operator $\mathcal P$ changes the sign of the momentum operator $\hat p$ and the position operator  $\hat x$: $\hat p\to-\hat p$  and  $\hat x\to-\hat x$. Time reversal $\mathcal T$ has the effect  $\hat p\to-\hat p$,  $\hat x\to \hat x$, and $i\to-i$. $\mathcal T$ changes the sign of $i$ because $\mathcal T$ is required to preserve the fundamental commutation relation $[\hat x,\hat p]=i\hbar$ \cite{Bender1998}.
For quantum mechanics in $S^1$, the parity operator $\mathcal P$ satisfies the following properties \cite{TanimuraS1}: $
\mathcal P^\dagger \mathcal P=\mathcal P^2=1$, $\mathcal P\hat W\mathcal P^{-1}=\hat W^\dagger$, $\mathcal P\hat\pi_\theta\mathcal P^{-1}=-\hat\pi_\theta$, and $\mathcal P\ket{l}=\ket{-l}$.
We show that the time reversal operator $\mathcal T$ shares the same properties. The time reversal operator for a bosonic particle is an anti-unitary operator which satisfies $\mathcal T^\dagger\mathcal T=\mathcal T^2=1$ \cite{Weinberg}.
Time reversal changes the direction of motion of a particle in $S^1$, i.e. $\hat\pi_\theta\mathcal T\ket{l}=-l\hbar\mathcal T\ket{l}$ should be satisfied. Therefore, we have $\mathcal T\ket{l}=a_l\ket{-l}$ with a coefficient $a_l$. Anti-unitarity of $\mathcal T$ implies $a_l=1$. Then, using Eq. \eqref{WAlgebragen} we readily obtain 
\begin{align*}
&\mathcal{P}^\dagger\mathcal{P}=\mathcal{P}^2=1,
\
\mathcal{T}^\dagger\mathcal{T}=\mathcal{T}^2=1,
\\
&\mathcal{P}\hat\pi_\theta\mathcal{P}^{-1}=-\hat\pi_\theta,
\
\mathcal{P}\hat W\mathcal{P}^{-1}=\hat W^\dagger,
\
\mathcal{P}\ket{l}=\ket{-l},
\\
&\mathcal{T}\hat\pi_\theta\mathcal{T}^{-1}=-\hat\pi_\theta,
\
\mathcal{T}\hat W\mathcal{T}^{-1}=\hat W^\dagger,
\
\mathcal{T}\ket{l}=\ket{-l}.
\end{align*}
So, $\hat \pi_\theta$, $\hat W$ and $\ket{l}$ are $\mathcal{PT}$-symmetric:
\begin{align*}
(\mathcal{PT})\hat\pi_\theta(\mathcal{PT})^{-1}=\hat\pi_\theta,
\\
(\mathcal{PT})\hat W(\mathcal{PT})^{-1}=\hat W,
\\
(\mathcal{PT})\ket{l}=\ket{l}.
\end{align*}
Note that the Hermitian time operators in  Eq. \eqref{TAB} and Eq. \eqref{TABonS1} change their signs under $\mathcal{PT}$, in agreement with their interpretation as quantization for the time variable $t$. (For quantum mechanics on $S^1$, $\mathcal{T}\hat W\mathcal{T}^{-1}=\hat W^\dagger$ is possible only if $\mathcal T$ changes the sign of  $i$.)

\subsection{$\mathcal{PT}$-symmetric time operator}
Now, we can define a $\mathcal{PT}$-symmetric time operator by
\begin{align*}
\hat T_{S^1}^{\mathcal{PT}}=\hat F-\hat{W}\hat\mu_1,
\end{align*}
where $\hat F$ is a real $\mathcal{PT}$-symmetric operator which commutes with $\hat H$. This operator satisfies the commutation relation
\begin{align}
[\hat H,\hat T_{S^1}^{\mathcal{PT}}]\ket{l}&=-\hbar\hat{W}\ket{l}.\label{TPTAlgebra}
\end{align}
Formally, we have
\begin{align*}
\hat T_{S^1}^\mathcal{PT}=i\hat x\hat p^{-1}+(\hat F-\hat\mu_1)+O(R^{-1}),
\end{align*}
so the matrix elements of $\hat T_{S^1}^\mathcal{PT}$ diverge as $R\to\infty$ unless we choose $\hat F$ carefully. The simplest form is $\hat F=\hat\mu_1$. Thus, we obtain the \textit{$\mathcal{PT}$-symmetric time operator in $S^1$}
\begin{align}
\hat T_{S^1}^{\mathcal{PT}}=(1-\hat W)\hat\mu_1=\hat T_{S^1}^\text{Re}+i\hat T_{S^1}.\label{ABonS1}
\end{align}
$\hat T_{S^1}$ and $\hat T_{S^1}^\text{Re}$ are defined by Eq. \eqref{TABonS1} and Eq. \eqref{TREonS1}, respectively. This $\mathcal{PT}$-symmetric operator is interpreted as an imaginary time operator $i\hat T_{S^1}$ (i.e. quantization of the imaginary time  $\tau=it$) with the addition of a quantum correction. $\mathcal{PT}$ changes the sign of $i$ as well as the sign of the time operator, hence an imaginary time operator is invariant under $\mathcal{PT}$. Conversely, it is also reasonable to define the time operator in Eq. \eqref{TABonS1} as the imaginary part of \eqref{ABonS1}. 
%The physical significance of $\hat T_{S^1}^\text{Re}$ can be understood by taking the large radius limit. Using $I=mR^2$, $l=Rk=Rp/\hbar$ and $\theta=x/R$, we find that $\hat T_{S^1}^\text{Re}$ has matrix elements
%\begin{align*}
%\hbox{$\braket{\theta|\hat T_{S^1}^\text{Re}|l}$}&\hbox{$=\frac{2I}{\hbar}\frac{(1-2l+2l \cos\theta-i \sin\theta)}{1-4l^2}\braket{\theta|l}$}
%\\
%&\hbox{$=-\frac{1}{4\pi}\frac{\lambda_\text{dB}}{v}\braket{\theta|l}+O(R^{-1}).$}
%\end{align*}
%In the second line we did a Taylor expansion for $\theta=x/R\ll 1$. Here $\lambda_\text{dB}=2\pi\hbar/p$ is the de Broglie wave and $v=p/m$ is the group velocity of the particle. The term $\lambda_\text{dB}/v$ describes a ``matter wave clock", i.e. because of the periodicity of a de Broglie wavelength, a particle moving with a fixed velocity v has an internal clock with period $\lambda_\text{dB}/v$ [28,29].

\begin{figure}
\includegraphics[width=0.5\linewidth]{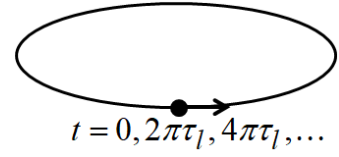}
\caption{The eigenvalues $\tau_l$ of $\hat T_{S^1}^{\mathcal{PT}}$ describes the periodicity in time of a free particle moving in a ring with a constant velocity.}
\label{ringsystem}
%\includegraphics[width=\linewidth]{plot.eps}
%\caption{The expectation value $\braket{\phi_l|\hat T_{S^1}(t)|\phi_l}/\braket{\phi_l|\phi_l}$ is shown for $l=0,1,5$ and $100$ for $\mu_{10}=2\times 10^{-6}$sec. The amplitude is proportional to $\mu_{1l}$ and the period is $P=2\pi\mu_{10}$.}
%\label{plot}
\end{figure}

The eigenstates and eigenvalues of $\hat T_{S^1}^{\mathcal{PT}}$ are calculated using biorthogonal quantum mechanics \cite{BOQM}: Suppose that the time operator $\hat T_{S^1}^{\mathcal{PT}}$ and its Hermitian conjugate $(\hat T_{S^1}^{\mathcal{PT}})^\dagger$ satisfy the eigenvalue equations
\begin{align}
\hat T_{S^1}^{\mathcal{PT}}\ket{\phi_l}=\tau_l\ket{\phi_l},\label{phieqn}
\\
(\hat T_{S^1}^{\mathcal{PT}})^\dagger\ket{\chi_l}=\tau_l^\ast\ket{\chi_l}.\label{chieqn}
\end{align}
Let us adopt the position representation $\hat\pi_\theta\to-i\hbar\frac{\partial}{\partial\theta}$ which implies $\hat\mu_1^{-1}\to\frac{\hbar}{I}\left(-i\frac{\partial}{\partial\theta}+\frac{1}{2}\right)$. Then, Eq. \eqref{phieqn} and Eq. \eqref{chieqn} are equivalent to the following differential equations:
\begin{align*}
\phi_l(\theta)&=\frac{\hbar}{I}\left(-i\frac{\partial}{\partial\theta}+\frac{1}{2}\right)\Phi_l(\theta),
\\
(1-e^{i\theta})\Phi_l(\theta)&=\frac{\tau_l\hbar}{I}\left(-i\frac{\partial}{\partial\theta}+\frac{1}{2}\right)\Phi_l(\theta),
\\
(1-e^{-i\theta})\chi_l(\theta)&=\frac{\tau_l^\ast\hbar}{I}\left(-i\frac{\partial}{\partial\theta}+\frac{1}{2}\right)\chi_l(\theta),
\end{align*}
which have the orthonormal solutions
\begin{align*}
\phi_l(\theta)&=\phi_{l0}\left(1-e^{i \theta }\right) e^{i \theta \nu_l}e^{-(\nu_l+\frac{1}{2})e^{i \theta }},
\\
\chi_l(\theta)&=\chi_{l0}e^{i \theta\nu_l^\ast}e^{(\nu_l^\ast+\frac{1}{2})e^{-i \theta }},
\\
\int_{-\pi}^\pi&d\theta \chi_l^\ast(\theta)\phi_m(\theta)=2\pi \chi^\ast_{l0}\phi_{m0}\delta_{l,m}=\delta_{l,m},
\end{align*}
where $\nu_l=\frac{I}{\tau_l\hbar}-\frac{1}{2}$ was introduced for brevity. The periodic boundary condition requires $\nu_l$ to be an integer. Therefore, we obtain the eigenvalues
\begin{equation}
\tau_l=\tau_l^\ast=\frac{2I}{(2\nu_l+1)\hbar}.\label{eigenvalue}
\end{equation}
These eigenvalues are interpreted as the time required for a free particle with velocity $v=(\nu_l+\frac{1}{2})\frac{\hbar}{mR}$ to move a distance $R$ on the ring; i.e. the time required to make a full rotation is $2\pi\tau_l=2\pi R/v$ (Fig. \ref{ringsystem}). The validity of this interpretation is supported by the generalized commutation relation. The Aharonov-Bohm time operator, for instance, is defined as an operator whose eigenvalue $t$ corresponds to the time required for a particle to move a finite distance $d$. Although $d/t$ gives the average velocity of the particle, the particle does not have a fixed momentum: Otherwise, the wave function of the particle is a momentum eigenstate $\psi_l$ and expectation value  with this state leads to contradictions (see Sec.\eqref{confinedsystem}). 
On the other hand, one can verify that the expectation value of Eq. \eqref{TPTAlgebra}  with biorthogonal quantum mechanics implies $\braket{\chi_l|[\hat H,\hat T_{S^1}^{\mathcal{PT}}]|\phi_l}=-\hbar\braket{\chi_l|\hat{W}|\phi_l}=0$. Therefore, our interpretation of time eigenvalues does not disagree with the commutation relation.
More generally, a wave function in the domain of $\hat T_{S^1}^{\mathcal{PT}}$ has the form $\ket{\Psi}=\sum_{l}a_l\ket{\phi_l}$ and the corresponding wave function in the domain of $(\hat T_{S^1}^{\mathcal{PT}})^\dagger$ has the form $\bra{\bar\Psi}=\sum_{l}a_l^\ast\bra{\chi_l}$. Arrival time in this state is 
\begin{equation}
\braket{\bar\Psi|\hat T_{S^1}^{\mathcal{PT}}|\Psi}=\sum_{l}|a_l|^2\tau_l.
\end{equation}
Commutation relations do not have a vanishing expectation value in this case. 

%Next, we calculate the large radius limit of $\hat T_{S^1}^\mathcal{PT}$. The eigenvalues $\tau_l$ diverge as $R\to\infty$ because it takes an infinite amount of time to move an infinite distance. Instead, if $\hat T_{S^1}^{\mathcal{PT}}$ has maximally broken $\mathcal{PT}$ symmetry; that is, if all of its eigenvalues $\tau_l$ are pure-imaginary complex-conjugate pairs, then $-i\hat T_{S^1}^{\mathcal{PT}}$(which is not defined in $\mathscr H$) has real eigenvalues. In this case, a calculation similar to Eq. \eqref{limit} gives
In the infinite-radius limit we obtain
\begin{align}
\hat T_{S^1}^\mathcal{PT}&\to i\hat T_\mathbb{R}^\text{NH}=i\hat T_\mathbb{R}-\frac{m\hbar}{2}\hat p^{-2}.
\label{PT_To_Recami}
\end{align}
As discussed after Eq. \eqref{TReAlgebra}, the last additional term describes a ``matter wave clock" which reflects the wave nature of a particle.
The non-Hermitian time operator $\hat T_\mathbb{R}^\text{NH}$ (which is not $\mathcal{PT}$-symmetric) was studied in ref. \cite{Recami10}. This operator satisfies the canonical commutation relation
\begin{equation}
[\hat H,\hat T_\mathbb{R}^\text{NH}]=[\hat H,(\hat T_\mathbb{R}^\text{NH})^\dagger]=[\hat H,\hat T_\mathbb{R}]=i\hbar.\label{TNRCCH}
\end{equation}
Note that $\hat T_{S^1}^{\mathcal{PT}}$ is also non-Hermitian for finite radius. We also note that other $\mathcal{PT}$-symmetric time operators can be defined naturally. $i\hat T_{\mathbb R}^\text{NH}$ is a non-Hermitian $\mathcal{PT}$-symmetric operator, but $\hat T_{\mathbb R}^\text{NH}$ is not $\mathcal{PT}$-symmetric. In addition, $i\hat T_{\mathbb R}$ is $\mathcal{PT}$-symmetric as well, but $i\hat T_{\mathbb R}$ and $\hat T_{\mathbb R}$ give the same time-energy uncertainty relation as we show in the next section.

%%%%%%%%%%%%%%%%%%%%%%%%%%
\section{Time-Energy Uncertainty Relations}
\label{TEUncertainty}
In the previous sections we saw that generalized commutation relations are necessary to define position operators and time operators in ring systems. Similarly, position-momentum uncertainty relations and time-energy uncertainty relations are modified in ring systems. This modification is motivated by the following example. Consider the uncertainty relations $\Delta \hat\theta\Delta \hat\pi_\theta\geq \hbar/2$ and $\Delta \hat H\Delta \hat T\geq\hbar/2$. For a momentum eigenstate we obtain $\Delta \hat\pi_\theta=0$ and $\Delta\hat H=0$ which imply the contradiction $0\geq\hbar/2$. This contradiction is avoided with generalized uncertainty relations.

The Robertson uncertainty relation \cite{RobertsonUR} between two Hermitian operators is given by 
\begin{equation}
(\Delta\hat A)_\psi(\Delta\hat B)_\psi\geq\frac{1}{2}\left|\langle[\hat A,\hat B]\rangle_\psi\right|,\label{RobertsonUR}
\end{equation}
where $\braket{\hat A}_\psi=\braket{\psi|\hat A|\psi}$ and $(\Delta\hat A)_\psi=\sqrt{\braket{\hat A^2}_\psi-\langle\hat A\rangle_\psi^2}$ is the norm of the state $[\hat A-\braket{\hat A}_\psi]\ket{\psi}$. For non-Hermitian operators we have a natural extension $(\Delta\hat A)_\psi=\sqrt{\braket{\hat A^\dagger\hat A}_\psi-|\braket{\hat A}_\psi|^2}$. In general, $(\Delta\hat A)_\psi\ne(\Delta\hat A^\dagger)_\psi$ if $\hat A$ and $\hat A^\dagger$ do not commute. Now, the Robertson uncertainty relation for non-Hermitian operators has a few different generalizations. First, Dou and Du \cite{DouDu} proposed the following generalization (which is written here with a different notation)
\begin{equation}
(\Delta\hat A)^0_\psi(\Delta\hat B)^0_\psi\geq\frac{1}{2}\left|\braket{[\hat A,\hat B]}_\psi\right|,\label{DouDuCCR}
\end{equation}
with $(\Delta\hat A)^0_\psi=[(\Delta\hat A)_\psi+(\Delta\hat A^\dagger)_\psi]/2$ and so on. Therefore, we have, for instance, the following uncertainty relations
\begin{align}
(\Delta\hat \pi_\theta)_\psi(\Delta\hat W)_\psi&\geq\frac{\hbar}{2}\left|\braket{\hat W}_\psi\right|,\label{DouDuW}
\\
(\Delta\hat H)_\psi(\Delta\hat T_{S^1}^{\mathcal{PT}})^0_\psi&\geq\frac{\hbar}{2}\left|\braket{\hat W}_\psi\right|.\label{DouDuT}
\end{align}
Second, Tanimura \cite{TanimuraS1} proposed the following uncertainty relation
\begin{align}
\begin{split}
(\Delta\hat\pi_\theta)_\psi(\Delta\hat W)_\psi&\geq\left|\braket{[\hat\pi_\theta-\braket{\hat \pi_\theta}_\psi][\hat W-\braket{\hat W}_\psi]}_\psi\right|
\\
&=\left|\braket{\hat\pi_\theta\hat W}_\psi-\braket{\hat \pi_\theta}_\psi\braket{\hat W}_\psi\right|,
\end{split}\label{TanimuraW}
\end{align}
which is a direct consequence of the Cauchy-Schwarz inequality. Similarly, we can obtain
\begin{align}
(\Delta\hat\pi_\theta)_\psi(\Delta\hat W)_\psi\geq\left|\braket{\hat \pi_\theta}_\psi\braket{\hat W}_\psi-\braket{\hat W\hat\pi_\theta}_\psi\right|,\label{NeW}
\\
(\Delta\hat H)_\psi(\Delta\hat T_{S^1}^{\mathcal{PT}})_\psi\geq|\braket{\hat H\hat T_{S^1}^{\mathcal{PT}}}_\psi-\braket{\hat H}_\psi\braket{\hat T_{S^1}^{\mathcal{PT}}}_\psi|,\label{T1}
\\
(\Delta\hat H)_\psi(\Delta\hat T_{S^1}^\mathcal{PT}{^\dagger})_\psi\geq|\braket{\hat H}_\psi\braket{\hat T_{S^1}^\mathcal{PT}}_\psi-\braket{\hat T_{S^1}^\mathcal{PT}\hat H}_\psi|.\label{T2}
\end{align}
We note that the uncertainty relations \eqref{TanimuraW}, \eqref{T1} and \eqref{T2} are not unique but other equivalent or inequivalent forms can be defined if necessary. Adding \eqref{T1} to \eqref{T2} and using the triangle inequality we obtain \eqref{DouDuT}. Moreover, by adding \eqref{TanimuraW} to the inequivalent form \eqref{NeW} we obtain \eqref{DouDuW}.
For the self-adjoint operators we have the usual Robertson uncertainty relations \eqref{RobertsonUR}. $\mathcal{PT}$-symmetric time operators satisfy Eq. \eqref{DouDuCCR}. In all cases, we see that the angle-angular momentum uncertainty relations and the energy-time uncertainty relations have the same lower bound. This fact is a consequence of the similarities between the commutation relations \eqref{TPTAlgebra}, \eqref{TReAlgebra}, \eqref{TImAlgebra}, \eqref{TThetaAlgebra} and \eqref{WAlgebragen}, \eqref{CSAlgebra}, \eqref{SCAlgebra}, \eqref{ThetaAlgebra} respectively. 

The significance of these uncertainty relations is as follows. 
First, uncertainties depend on wave functions and, on the contrary, the state of a ring system can be inferred from uncertainties. Under some circumstances, such as for non-localized momentum eigenstates, we have $|\braket{[\hat H,\hat T]}|=0$, i.e. uncertainty vanishes. We have $|\braket{[\hat H,\hat T]}|=\hbar$ for a well-localized state or for time operators introduced by Galapon and collaborators \cite{Galapon451, GalaponDiscrete}, which reduces to the case of canonical commutation relation. In fact, the same situation appears for angle-angular momentum commutation relations as studied in ref. \cite{TanimuraS1}. Second, we cannot completely distinguish from the canonical commutation relation and from the time-energy uncertainty relation if a time operator is self-adjoint or $\mathcal{PT}$ symmetric: For instance, the Aharonov-Bohm time operator defined in Eq. \eqref{TAB} and the non-Hermitian time operator in Eq. \eqref{PT_To_Recami} give the same commutation relation (Eq. \eqref{TNRCCH}) and the same time-energy uncertainty relation (as it can be seen from Eq. \eqref{TNRCCH}, Eq. \eqref{RobertsonUR}, Eq. \eqref{DouDuCCR}). Therefore, if time operators are defined based on uncertainty relations, then it is possible that such time operators are not self-adjoint (not even Hermitian) but $\mathcal{PT}$-symmetric. On the other hand, their distinction is possible for our time operators, so it is important to predict possible experimental uncertainties.

\section{Discussion and Conclusion}
\begin{figure}
\includegraphics[width=0.75\linewidth]{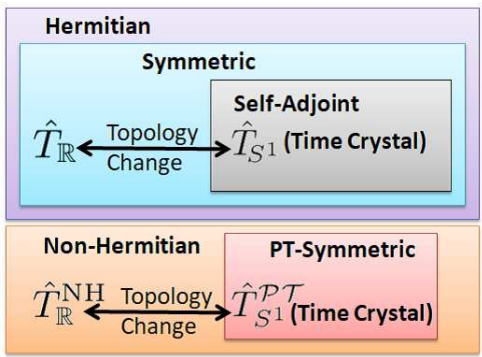}
\caption{$\hat T_{S^1}$ [Eq. \eqref{TABonS1}] is a self-adjoint time operator and $\hat T_{S^1}^{\mathcal{PT}}$ [Eq. \eqref{ABonS1}] is a $\mathcal{PT}$-symmetric time operator. In the large radius limit (from $S^1$ to $\mathbb{R}$), $\hat T_{S^1}$ reduces to the Aharonov-Bohm time operator $\hat T_\mathbb{R}$ [Eq. \eqref{TAB}] and $\hat T_{S^1}^{\mathcal{PT}}$ reduces to the non-Hermitian operator $\hat T_\mathbb{R}^\text{NH}$ [Eq. \eqref{PT_To_Recami}] \cite{note2}.}
\label{Inclusion_relation}
\end{figure}

%%%%%%%%%%%%%%%%%%%%%%%%%%
First, we summarize our main results. We defined a self-adjoint time operator $\hat T$ and a $\mathcal{PT}$-symmetric time operator $\hat T_{S^1}^\mathcal{PT}$ for a free particle on a ring system. $\hat T$ is a generalized time operator which satisfies Eq. \eqref{GWWRDef}. We studied three cases of $\hat T$, namely 1) $\hat T_{S^1}$ which can be used for our previous model of QTC, 2) $\hat T_{S^1}^\text{Re}$ which can be interpreted as a ``matter wave clock", and $\hat T_\Theta$ which is interpreted as a time-of-arrival operator and contains Galapon's time operator for periodic systems as a special case. The eigenstates and eigenvalues of $\hat T_{S^1}^{\mathcal{PT}}$ are calculated using biorthogonal quantum mechanics {\cite{BOQM}}. A summary of various time operators is given in Fig. \ref{Inclusion_relation}.

Second, Galapon \textit{et al.} also defined a self-adjoint time operator for a confined system with and without periodic boundary conditions which satisfy Eq. \eqref{CCR1} \cite{Galapon451,CQTA}. However, our results suggest that time operators should be defined based on the real-space topology of a quantum system. So, Eq. \eqref{CCR1} may not be the only possibility to define self-adjoint time operators. Our work may also be generalized to relativistic particles, but quantization of constrained systems is still a subject of current active research and the commutation relations Eq. \eqref{WAlgebragen} are not guaranteed to hold for relativistic systems as well.

Third, in order to obtain additional insights into QTC and understand how to create many other models of QTC, a universal origin of this periodicity is required. We surmise that this origin can be understood from Eq. \eqref{GWWRDef} and $\hat K'(t)$ characterizes the time-periodic evolution of a QTC: For our previous QTC model, this quantity is the charge density expectation value; and for Floquet time crystals \cite{Sacha,FTC, PhaseStructure,FTCTheory, DTCYao, Prethermal, FTCChoi, FTCZhang} this quantity is typically the single-ion magnetization. In addition, we note that most of the early models of QTC \cite{QTC, Wigner, WilczekFFLO, Nozieres} are ring systems but the infinite-radius approximation was used. Therefore, these models should be reconsidered in light of our results.

Fourth, an operator being $\mathcal{PT}$-symmetric does not necessarily mean that it has real eigenvalues, but it can have real eigenvalues.  Typically, in $\mathcal{PT}$-symmetric systems with gain and loss (denoted by $\pm\gamma$) all eigenvalues are real below a critical value $\gamma < \gamma_c$ and $\mathcal{PT}$-symmetry is spontaneously broken otherwise.  Therefore, it would be an interesting problem to consider the spontaneous breaking of $\mathcal{PT}$-symmetry (of the Hamiltonian or the time operator) in periodically driven Floquet time crystals \cite{Sacha,FTC, PhaseStructure,FTCTheory, DTCYao, Prethermal, FTCChoi, FTCZhang}.

Moreover, possible applications of work are as follows. 1) As we discussed in Sec. \ref{TC}, the GWWR may give a universal explanation of time crystals \cite{QTC,Wigner, WilczekFFLO,Nozieres,Sacha,FTC, PhaseStructure,
FTCTheory,DTCYao,Prethermal,
FTCChoi,FTCZhang,CDWQTC}, where the commutation factor $\hat K(t)=\hat T(t)-\hat T$ gives the temporal structure of the system. 2) Our time operator is defined based on the position operator $\hat f$. Therefore, our work may be used to reexamine the relationship between space and time in the level of operators and understand space-time structures in quantum mechanics. 3) More radically, analysis of the cosmic microwave background anisotropy suggests that our universe is finite with a periodic structure \cite{Luminet2003}. If this is correct, then our work may be used in quantum cosmology to understand, for instance, the space-time structure of the early (topologically non-trivial) universe.

Finally, we discuss the results in this paper in the context of experiments. 
Suppose we want to study experimentally the arrival time or other events of a ring system. According to Sec. \ref{confinedsystem}, time operators with the CCR can be studied only for special kinds of states and experiments must be well-designed. On the other hand, our time operators can be used to study time of arrival or other events for general states in ring systems. For example, the order parameter of a charge density wave (CDW) or a superconductor is a complex scalar $\Delta=|\Delta|e^{i\theta}$. Suppose that we can quantize the phase $\theta$ by $\hat W=\widehat{e^{i\theta}}$. Then the commutation relations Eq. \eqref{TImAlgebra}, Eq. \eqref{TReAlgebra}, and Eq. \eqref{TPTAlgebra} give the Heisenberg equations of motion of the time operators $\frac{d}{dt}\hat T_{S^1}(t)=-\hat{C}(t)$, $\frac{d}{dt}\hat T_{S^1}^\text{Re}(t)=\hat{S}(t)$, and $\frac{d}{dt}\hat T_{S^1}^\mathcal{PT}(t)=-i\hat W(t)$, respectively. $\hat C(t)$ can describe charge density fluctuation and $\hat S(t)$ can describe time-dependent Josephson current. Therefore, our results can be tested in various CDW \cite{ring,CDWQTC} or superconducting systems (including topological crystals such as ring crystals and M\"obius strip crystals \cite{ring,Povie172}), in superfluid systems with spatial periodicity \cite{Nozieres,WilczekFFLO} and in ring-shaped ion traps \cite{Wigner,Haffner}. $\hat T_\Theta$ can be tested by measuring the time-of-arrival of a sine-Gordon-soliton in annular Josephson junctions \cite{AnnularSoliton}, or we can use any other systems which can be described by a single particle in $S^1$. Experimental uncertainties will be given by the results in Sec. \ref{TEUncertainty}. Uncertainties depend on wave functions and, on the contrary, the state of a ring system can be inferred from uncertainties. Furthermore, uncertainties themselves can be used to test our formulation in experiments.
Experimental scheme to study uncertainty relations is possible and has been used in a different context (such as \cite{Hasegawa}). We believe that similar experiments are possible for ring systems.

\acknowledgments
We thank  Asao Arai, Izumi Tsutsui, Yoshimasa Hidaka, Akio Hosoya, Yuji Hasegawa, Toyoki Matsuyama, Kohkichi Konno, Kousuke Yakubo, Hideaki Obuse and Koichi Ichimura for stimulating and valuable discussions. This work was supported in part by the U. S. Department of Energy.

\onecolumngrid
\bibliographystyle{apsrev4-1}
\bibliography{20190324_Ver13Time_Operators_and_Time_Crystals}

%merlin.mbs apsrev4-1.bst 2010-07-25 4.21a (PWD, AO, DPC) hacked
%Control: key (0)
%Control: author (72) initials jnrlst
%Control: editor formatted (1) identically to author
%Control: production of article title (-1) disabled
%Control: page (0) single
%Control: year (1) truncated
%Control: production of eprint (0) enabled
\begin{thebibliography}{48}%
\makeatletter
\providecommand \@ifxundefined [1]{%
 \@ifx{#1\undefined}
}%
\providecommand \@ifnum [1]{%
 \ifnum #1\expandafter \@firstoftwo
 \else \expandafter \@secondoftwo
 \fi
}%
\providecommand \@ifx [1]{%
 \ifx #1\expandafter \@firstoftwo
 \else \expandafter \@secondoftwo
 \fi
}%
\providecommand \natexlab [1]{#1}%
\providecommand \enquote  [1]{``#1''}%
\providecommand \bibnamefont  [1]{#1}%
\providecommand \bibfnamefont [1]{#1}%
\providecommand \citenamefont [1]{#1}%
\providecommand \href@noop [0]{\@secondoftwo}%
\providecommand \href [0]{\begingroup \@sanitize@url \@href}%
\providecommand \@href[1]{\@@startlink{#1}\@@href}%
\providecommand \@@href[1]{\endgroup#1\@@endlink}%
\providecommand \@sanitize@url [0]{\catcode `\\12\catcode `\$12\catcode
  `\&12\catcode `\#12\catcode `\^12\catcode `\_12\catcode `\%12\relax}%
\providecommand \@@startlink[1]{}%
\providecommand \@@endlink[0]{}%
\providecommand \url  [0]{\begingroup\@sanitize@url \@url }%
\providecommand \@url [1]{\endgroup\@href {#1}{\urlprefix }}%
\providecommand \urlprefix  [0]{URL }%
\providecommand \Eprint [0]{\href }%
\providecommand \doibase [0]{http://dx.doi.org/}%
\providecommand \selectlanguage [0]{\@gobble}%
\providecommand \bibinfo  [0]{\@secondoftwo}%
\providecommand \bibfield  [0]{\@secondoftwo}%
\providecommand \translation [1]{[#1]}%
\providecommand \BibitemOpen [0]{}%
\providecommand \bibitemStop [0]{}%
\providecommand \bibitemNoStop [0]{.\EOS\space}%
\providecommand \EOS [0]{\spacefactor3000\relax}%
\providecommand \BibitemShut  [1]{\csname bibitem#1\endcsname}%
\let\auto@bib@innerbib\@empty
%</preamble>
\bibitem [{More precisely()}]{note1}%
  \BibitemOpen
  More precisely,\ \href@noop {} {}\bibinfo {note} {{if the eigenvalues of a
  Hamiltonian are continuous and bounded below, then there is no self-adjoint
  time operators which satisfy the canonical commutation relation
  (1).}}\BibitemShut {Stop}%
\bibitem [{\citenamefont {Reed}\ and\ \citenamefont
  {Simon}(1975)}]{ReedAndSimon}%
  \BibitemOpen
  \bibfield  {author} {\bibinfo {author} {\bibfnamefont {M.}~\bibnamefont
  {Reed}}\ and\ \bibinfo {author} {\bibfnamefont {B.}~\bibnamefont {Simon}},\
  }\href
  {https://www.amazon.com/Fourier-Analysis-Self-Adjointness-Methods-Mathematical/dp/0125850026?SubscriptionId=0JYN1NVW651KCA56C102&tag=techkie-20&linkCode=xm2&camp=2025&creative=165953&creativeASIN=0125850026}
  {\emph {\bibinfo {title} {Fourier Analysis, Self-Adjointness (Methods of
  Modern Mathematical Physics, Vol. 2)}}}\ (\bibinfo  {publisher} {Academic
  Press},\ \bibinfo {year} {1975})\BibitemShut {NoStop}%
\bibitem [{\citenamefont {Muga}\ and\ \citenamefont
  {Leavens}(2000)}]{Arrival_Time_in_Quantum_Mechanics}%
  \BibitemOpen
  \bibfield  {author} {\bibinfo {author} {\bibfnamefont {J.}~\bibnamefont
  {Muga}}\ and\ \bibinfo {author} {\bibfnamefont {C.}~\bibnamefont {Leavens}},\
  }\href {\doibase https://doi.org/10.1016/S0370-1573(00)00047-8} {\bibfield
  {journal} {\bibinfo  {journal} {Phys. Rep.}\ }\textbf {\bibinfo {volume}
  {338}},\ \bibinfo {pages} {353 } (\bibinfo {year} {2000})}\BibitemShut
  {NoStop}%
\bibitem [{\citenamefont {Aharonov}\ and\ \citenamefont
  {Bohm}(1961)}]{ABTimeOperator}%
  \BibitemOpen
  \bibfield  {author} {\bibinfo {author} {\bibfnamefont {Y.}~\bibnamefont
  {Aharonov}}\ and\ \bibinfo {author} {\bibfnamefont {D.}~\bibnamefont
  {Bohm}},\ }\href {\doibase 10.1103/PhysRev.122.1649} {\bibfield  {journal}
  {\bibinfo  {journal} {Phys. Rev.}\ }\textbf {\bibinfo {volume} {122}},\
  \bibinfo {pages} {1649} (\bibinfo {year} {1961})}\BibitemShut {NoStop}%
\bibitem [{\citenamefont {Pauli}()}]{Pauli1958}%
  \BibitemOpen
  \bibfield  {author} {\bibinfo {author} {\bibfnamefont {W.}~\bibnamefont
  {Pauli}},\ }in\ \href@noop {} {\emph {\bibinfo {booktitle} {Handbuch der
  Physik (Encyclopedia of physics)}}},\ \bibinfo {editor} {edited by\ \bibinfo
  {editor} {\bibfnamefont {S.}~\bibnamefont {Fludge}}},\ pp.\ \bibinfo {pages}
  {1--168}\BibitemShut {NoStop}%
\bibitem [{\citenamefont {Galapon}(2002{\natexlab{a}})}]{Galapon451}%
  \BibitemOpen
  \bibfield  {author} {\bibinfo {author} {\bibfnamefont {E.}~\bibnamefont
  {Galapon}},\ }\href {\doibase 10.1098/rspa.2001.0874} {\bibfield  {journal}
  {\bibinfo  {journal} {Proc. Royal Soc. Lond. A: Math. Phys. Eng. Sci.}\
  }\textbf {\bibinfo {volume} {458}},\ \bibinfo {pages} {451} (\bibinfo {year}
  {2002}{\natexlab{a}})}\BibitemShut {NoStop}%
\bibitem [{\citenamefont {Galapon}\ \emph {et~al.}(2004)\citenamefont
  {Galapon}, \citenamefont {Caballar},\ and\ \citenamefont
  {Bahague~Jr}}]{CQTA}%
  \BibitemOpen
  \bibfield  {author} {\bibinfo {author} {\bibfnamefont {E.~A.}\ \bibnamefont
  {Galapon}}, \bibinfo {author} {\bibfnamefont {R.~F.}\ \bibnamefont
  {Caballar}}, \ and\ \bibinfo {author} {\bibfnamefont {R.~T.}\ \bibnamefont
  {Bahague~Jr}},\ }\href {\doibase 10.1103/PhysRevLett.93.180406} {\bibfield
  {journal} {\bibinfo  {journal} {Phys. Rev. Lett.}\ }\textbf {\bibinfo
  {volume} {93}},\ \bibinfo {pages} {180406} (\bibinfo {year}
  {2004})}\BibitemShut {NoStop}%
\bibitem [{\citenamefont {Galapon}(2002{\natexlab{b}})}]{GalaponDiscrete}%
  \BibitemOpen
  \bibfield  {author} {\bibinfo {author} {\bibfnamefont {E.~A.}\ \bibnamefont
  {Galapon}},\ }\href@noop {} {\bibfield  {journal} {\bibinfo  {journal} {Proc.
  Royal Soc. Lond. A: Math. Phys. Eng. Sci.}\ }\textbf {\bibinfo {volume}
  {458}},\ \bibinfo {pages} {2671} (\bibinfo {year}
  {2002}{\natexlab{b}})}\BibitemShut {NoStop}%
\bibitem [{\citenamefont {Arai}\ and\ \citenamefont
  {Matsuzawa}(2008)}]{AraiMatsuzawa}%
  \BibitemOpen
  \bibfield  {author} {\bibinfo {author} {\bibfnamefont {A.}~\bibnamefont
  {Arai}}\ and\ \bibinfo {author} {\bibfnamefont {Y.}~\bibnamefont
  {Matsuzawa}},\ }\href@noop {} {\bibfield  {journal} {\bibinfo  {journal}
  {Rev. Math. Phys.}\ }\textbf {\bibinfo {volume} {20}},\ \bibinfo {pages}
  {951} (\bibinfo {year} {2008})}\BibitemShut {NoStop}%
\bibitem [{\citenamefont {Arai}(2008)}]{AraiDiscrete}%
  \BibitemOpen
  \bibfield  {author} {\bibinfo {author} {\bibfnamefont {A.}~\bibnamefont
  {Arai}},\ }\href@noop {} {\bibfield  {journal} {\bibinfo  {journal} {Lett.
  Math. Phys.}\ }\textbf {\bibinfo {volume} {87}},\ \bibinfo {pages} {67}
  (\bibinfo {year} {2008})}\BibitemShut {NoStop}%
\bibitem [{\citenamefont {Wilczek}(2012)}]{QTC}%
  \BibitemOpen
  \bibfield  {author} {\bibinfo {author} {\bibfnamefont {F.}~\bibnamefont
  {Wilczek}},\ }\href@noop {} {\bibfield  {journal} {\bibinfo  {journal} {Phys.
  Rev. Lett.}\ }\textbf {\bibinfo {volume} {{\bf 109}}},\ \bibinfo {pages}
  {160401} (\bibinfo {year} {2012})}\BibitemShut {NoStop}%
\bibitem [{\citenamefont {Li}\ \emph {et~al.}(2012)\citenamefont {Li} \emph
  {et~al.}}]{Wigner}%
  \BibitemOpen
  \bibfield  {author} {\bibinfo {author} {\bibfnamefont {T.}~\bibnamefont {Li}}
  \emph {et~al.},\ }\href {\doibase 10.1103/PhysRevLett.109.163001} {\bibfield
  {journal} {\bibinfo  {journal} {Phys. Rev. Lett.}\ }\textbf {\bibinfo
  {volume} {109}},\ \bibinfo {pages} {163001} (\bibinfo {year}
  {2012})}\BibitemShut {NoStop}%
\bibitem [{\citenamefont {Wilczek}(2013)}]{WilczekFFLO}%
  \BibitemOpen
  \bibfield  {author} {\bibinfo {author} {\bibfnamefont {F.}~\bibnamefont
  {Wilczek}},\ }\href {\doibase 10.1103/PhysRevLett.111.250402} {\bibfield
  {journal} {\bibinfo  {journal} {Phys. Rev. Lett.}\ }\textbf {\bibinfo
  {volume} {111}},\ \bibinfo {pages} {250402} (\bibinfo {year}
  {2013})}\BibitemShut {NoStop}%
\bibitem [{\citenamefont {Nakatsugawa}\ \emph {et~al.}(2017)\citenamefont
  {Nakatsugawa}, \citenamefont {Fujii},\ and\ \citenamefont {Tanda}}]{CDWQTC}%
  \BibitemOpen
  \bibfield  {author} {\bibinfo {author} {\bibfnamefont {K.}~\bibnamefont
  {Nakatsugawa}}, \bibinfo {author} {\bibfnamefont {T.}~\bibnamefont {Fujii}},
  \ and\ \bibinfo {author} {\bibfnamefont {S.}~\bibnamefont {Tanda}},\ }\href
  {\doibase 10.1103/PhysRevB.96.094308} {\bibfield  {journal} {\bibinfo
  {journal} {Phys. Rev. B}\ }\textbf {\bibinfo {volume} {96}},\ \bibinfo
  {pages} {094308} (\bibinfo {year} {2017})}\BibitemShut {NoStop}%
\bibitem [{\citenamefont {Nozi\`{e}res}(2013)}]{Nozieres}%
  \BibitemOpen
  \bibfield  {author} {\bibinfo {author} {\bibfnamefont {P.}~\bibnamefont
  {Nozi\`{e}res}},\ }\href {http://stacks.iop.org/0295-5075/103/i=5/a=57008}
  {\bibfield  {journal} {\bibinfo  {journal} {Europhys. Lett.}\ }\textbf
  {\bibinfo {volume} {103}},\ \bibinfo {pages} {57008} (\bibinfo {year}
  {2013})}\BibitemShut {NoStop}%
\bibitem [{\citenamefont {Sacha}(2015)}]{Sacha}%
  \BibitemOpen
  \bibfield  {author} {\bibinfo {author} {\bibfnamefont {K.}~\bibnamefont
  {Sacha}},\ }\href {\doibase 10.1103/PhysRevA.91.033617} {\bibfield  {journal}
  {\bibinfo  {journal} {Phys. Rev. A}\ }\textbf {\bibinfo {volume} {91}},\
  \bibinfo {pages} {033617} (\bibinfo {year} {2015})}\BibitemShut {NoStop}%
\bibitem [{\citenamefont {Else}\ \emph {et~al.}(2016)\citenamefont {Else},
  \citenamefont {Bauer},\ and\ \citenamefont {Nayak}}]{FTC}%
  \BibitemOpen
  \bibfield  {author} {\bibinfo {author} {\bibfnamefont {D.~V.}\ \bibnamefont
  {Else}}, \bibinfo {author} {\bibfnamefont {B.}~\bibnamefont {Bauer}}, \ and\
  \bibinfo {author} {\bibfnamefont {C.}~\bibnamefont {Nayak}},\ }\href
  {\doibase 10.1103/PhysRevLett.117.090402} {\bibfield  {journal} {\bibinfo
  {journal} {Phys. Rev. Lett.}\ }\textbf {\bibinfo {volume} {117}},\ \bibinfo
  {pages} {090402} (\bibinfo {year} {2016})}\BibitemShut {NoStop}%
\bibitem [{\citenamefont {Khemani}\ \emph {et~al.}(2016)\citenamefont
  {Khemani}, \citenamefont {Lazarides}, \citenamefont {Moessner},\ and\
  \citenamefont {Sondhi}}]{PhaseStructure}%
  \BibitemOpen
  \bibfield  {author} {\bibinfo {author} {\bibfnamefont {V.}~\bibnamefont
  {Khemani}}, \bibinfo {author} {\bibfnamefont {A.}~\bibnamefont {Lazarides}},
  \bibinfo {author} {\bibfnamefont {R.}~\bibnamefont {Moessner}}, \ and\
  \bibinfo {author} {\bibfnamefont {S.~L.}\ \bibnamefont {Sondhi}},\ }\href
  {\doibase 10.1103/PhysRevLett.116.250401} {\bibfield  {journal} {\bibinfo
  {journal} {Phys. Rev. Lett.}\ }\textbf {\bibinfo {volume} {116}},\ \bibinfo
  {pages} {250401} (\bibinfo {year} {2016})}\BibitemShut {NoStop}%
\bibitem [{\citenamefont {von Keyserlingk}\ and\ \citenamefont
  {Sondhi}(2016)}]{FTCTheory}%
  \BibitemOpen
  \bibfield  {author} {\bibinfo {author} {\bibfnamefont {C.~W.}\ \bibnamefont
  {von Keyserlingk}}\ and\ \bibinfo {author} {\bibfnamefont {S.~L.}\
  \bibnamefont {Sondhi}},\ }\href@noop {} {\bibfield  {journal} {\bibinfo
  {journal} {Phys. Rev. B}\ }\textbf {\bibinfo {volume} {{\bf 93}}},\ \bibinfo
  {pages} {245146} (\bibinfo {year} {2016})}\BibitemShut {NoStop}%
\bibitem [{\citenamefont {Yao}\ \emph {et~al.}(2017)\citenamefont {Yao},
  \citenamefont {Potter}, \citenamefont {Potirniche},\ and\ \citenamefont
  {Vishwanath}}]{DTCYao}%
  \BibitemOpen
  \bibfield  {author} {\bibinfo {author} {\bibfnamefont {N.~Y.}\ \bibnamefont
  {Yao}}, \bibinfo {author} {\bibfnamefont {A.~C.}\ \bibnamefont {Potter}},
  \bibinfo {author} {\bibfnamefont {I.-D.}\ \bibnamefont {Potirniche}}, \ and\
  \bibinfo {author} {\bibfnamefont {A.}~\bibnamefont {Vishwanath}},\
  }\href@noop {} {\bibfield  {journal} {\bibinfo  {journal} {Phys. Rev. Lett.}\
  }\textbf {\bibinfo {volume} {{\bf 118}}},\ \bibinfo {pages} {030401}
  (\bibinfo {year} {2017})}\BibitemShut {NoStop}%
\bibitem [{\citenamefont {Else}\ \emph {et~al.}(2017)\citenamefont {Else},
  \citenamefont {Bauer},\ and\ \citenamefont {Nayak}}]{Prethermal}%
  \BibitemOpen
  \bibfield  {author} {\bibinfo {author} {\bibfnamefont {D.~V.}\ \bibnamefont
  {Else}}, \bibinfo {author} {\bibfnamefont {B.}~\bibnamefont {Bauer}}, \ and\
  \bibinfo {author} {\bibfnamefont {C.}~\bibnamefont {Nayak}},\ }\href@noop {}
  {\bibfield  {journal} {\bibinfo  {journal} {Phys. Rev. X}\ }\textbf {\bibinfo
  {volume} {{\bf 7}}},\ \bibinfo {pages} {011026} (\bibinfo {year}
  {2017})}\BibitemShut {NoStop}%
\bibitem [{\citenamefont {Choi}\ \emph {et~al.}(2017)\citenamefont {Choi} \emph
  {et~al.}}]{FTCChoi}%
  \BibitemOpen
  \bibfield  {author} {\bibinfo {author} {\bibfnamefont {S.}~\bibnamefont
  {Choi}} \emph {et~al.},\ }\href@noop {} {\bibfield  {journal} {\bibinfo
  {journal} {Nature}\ }\textbf {\bibinfo {volume} {{\bf 543}}},\ \bibinfo
  {pages} {9} (\bibinfo {year} {2017})}\BibitemShut {NoStop}%
\bibitem [{\citenamefont {Zhang}\ \emph {et~al.}(2017)\citenamefont {Zhang}
  \emph {et~al.}}]{FTCZhang}%
  \BibitemOpen
  \bibfield  {author} {\bibinfo {author} {\bibfnamefont {J.}~\bibnamefont
  {Zhang}} \emph {et~al.},\ }\href@noop {} {\bibfield  {journal} {\bibinfo
  {journal} {Nature}\ }\textbf {\bibinfo {volume} {{\bf 543}}},\ \bibinfo
  {pages} {217} (\bibinfo {year} {2017})}\BibitemShut {NoStop}%
\bibitem [{\citenamefont {Arai}(2005)}]{GWWR}%
  \BibitemOpen
  \bibfield  {author} {\bibinfo {author} {\bibfnamefont {A.}~\bibnamefont
  {Arai}},\ }\href@noop {} {\bibfield  {journal} {\bibinfo  {journal} {Rev.
  Math. Phys.}\ }\textbf {\bibinfo {volume} {17}},\ \bibinfo {pages} {1071}
  (\bibinfo {year} {2005})}\BibitemShut {NoStop}%
\bibitem [{\citenamefont {Bender}\ \emph {et~al.}(1999)\citenamefont {Bender},
  \citenamefont {Boettcher},\ and\ \citenamefont {Meisinger}}]{Bender1998}%
  \BibitemOpen
  \bibfield  {author} {\bibinfo {author} {\bibfnamefont {C.~M.}\ \bibnamefont
  {Bender}}, \bibinfo {author} {\bibfnamefont {S.}~\bibnamefont {Boettcher}}, \
  and\ \bibinfo {author} {\bibfnamefont {P.~N.}\ \bibnamefont {Meisinger}},\
  }\href {\doibase 10.1063/1.532860} {\bibfield  {journal} {\bibinfo  {journal}
  {J. Math. Phys.}\ }\textbf {\bibinfo {volume} {40}},\ \bibinfo {pages} {2201}
  (\bibinfo {year} {1999})}\BibitemShut {NoStop}%
\bibitem [{\citenamefont {Bender}(2007)}]{Bender2007}%
  \BibitemOpen
  \bibfield  {author} {\bibinfo {author} {\bibfnamefont {C.~M.}\ \bibnamefont
  {Bender}},\ }\href {http://stacks.iop.org/0034-4885/70/i=6/a=R03} {\bibfield
  {journal} {\bibinfo  {journal} {Rep. Prog. Phys.}\ }\textbf {\bibinfo
  {volume} {70}},\ \bibinfo {pages} {947} (\bibinfo {year} {2007})}\BibitemShut
  {NoStop}%
\bibitem [{\citenamefont {Bender}(2015)}]{Bender2015}%
  \BibitemOpen
  \bibfield  {author} {\bibinfo {author} {\bibfnamefont {C.~M.}\ \bibnamefont
  {Bender}},\ }\href@noop {} {\bibfield  {journal} {\bibinfo  {journal}
  {Journal of Physics: Conference Series}\ }\textbf {\bibinfo {volume} {631}},\
  \bibinfo {pages} {012002} (\bibinfo {year} {2015})}\BibitemShut {NoStop}%
\bibitem [{\citenamefont {Recami}\ \emph {et~al.}(2010)\citenamefont {Recami},
  \citenamefont {Olkhovsky},\ and\ \citenamefont {Maydanyuk}}]{Recami10}%
  \BibitemOpen
  \bibfield  {author} {\bibinfo {author} {\bibfnamefont {E.}~\bibnamefont
  {Recami}}, \bibinfo {author} {\bibfnamefont {V.~S.}\ \bibnamefont
  {Olkhovsky}}, \ and\ \bibinfo {author} {\bibfnamefont {S.~P.}\ \bibnamefont
  {Maydanyuk}},\ }\href {\doibase 10.1142/S0217751X10048007} {\bibfield
  {journal} {\bibinfo  {journal} {Int. J. Mod. Phys. A}\ }\textbf {\bibinfo
  {volume} {25}},\ \bibinfo {pages} {1785} (\bibinfo {year}
  {2010})}\BibitemShut {NoStop}%
\bibitem [{\citenamefont {Schm{\"u}dgen}(1983)}]{WWE}%
  \BibitemOpen
  \bibfield  {author} {\bibinfo {author} {\bibfnamefont {K.}~\bibnamefont
  {Schm{\"u}dgen}},\ }\href {\doibase
  https://doi.org/10.1016/0022-1236(83)90058-7} {\bibfield  {journal} {\bibinfo
   {journal} {J. Funct. Anal.}\ }\textbf {\bibinfo {volume} {50}},\ \bibinfo
  {pages} {8 } (\bibinfo {year} {1983})}\BibitemShut {NoStop}%
\bibitem [{\citenamefont {Miyamoto}(2001)}]{Miyamoto}%
  \BibitemOpen
  \bibfield  {author} {\bibinfo {author} {\bibfnamefont {M.}~\bibnamefont
  {Miyamoto}},\ }\href {\doibase 10.1063/1.1346598} {\bibfield  {journal}
  {\bibinfo  {journal} {J. Math. Phys.}\ }\textbf {\bibinfo {volume} {42}},\
  \bibinfo {pages} {1038} (\bibinfo {year} {2001})}\BibitemShut {NoStop}%
\bibitem [{\citenamefont {Arai}\ and\ \citenamefont
  {Hiroshima}(2017)}]{AraiUWTO}%
  \BibitemOpen
  \bibfield  {author} {\bibinfo {author} {\bibfnamefont {A.}~\bibnamefont
  {Arai}}\ and\ \bibinfo {author} {\bibfnamefont {F.}~\bibnamefont
  {Hiroshima}},\ }\href@noop {} {\bibfield  {journal} {\bibinfo  {journal}
  {Ann. Henri Poincar{\'e}}\ }\textbf {\bibinfo {volume} {18}},\ \bibinfo
  {pages} {2995} (\bibinfo {year} {2017})}\BibitemShut {NoStop}%
\bibitem [{\citenamefont {Ohnuki}\ and\ \citenamefont
  {Kitakado}(1993)}]{Ohnuki}%
  \BibitemOpen
  \bibfield  {author} {\bibinfo {author} {\bibfnamefont {Y.}~\bibnamefont
  {Ohnuki}}\ and\ \bibinfo {author} {\bibfnamefont {S.}~\bibnamefont
  {Kitakado}},\ }\href@noop {} {\bibfield  {journal} {\bibinfo  {journal} {J.
  Math. Phys.}\ }\textbf {\bibinfo {volume} {{\bf 34}}},\ \bibinfo {pages}
  {2827} (\bibinfo {year} {1993})}\BibitemShut {NoStop}%
\bibitem [{\citenamefont {Carruthers}\ and\ \citenamefont
  {Nieto}(1968)}]{CARRUTHERSandNIETO}%
  \BibitemOpen
  \bibfield  {author} {\bibinfo {author} {\bibfnamefont {P.}~\bibnamefont
  {Carruthers}}\ and\ \bibinfo {author} {\bibfnamefont {M.~M.}\ \bibnamefont
  {Nieto}},\ }\href {\doibase 10.1103/RevModPhys.40.411} {\bibfield  {journal}
  {\bibinfo  {journal} {Rev. Mod. Phys.}\ }\textbf {\bibinfo {volume} {40}},\
  \bibinfo {pages} {411} (\bibinfo {year} {1968})}\BibitemShut {NoStop}%
\bibitem [{\citenamefont {Tanimura}(1993)}]{TanimuraS1}%
  \BibitemOpen
  \bibfield  {author} {\bibinfo {author} {\bibfnamefont {S.}~\bibnamefont
  {Tanimura}},\ }\href {\doibase 10.1143/ptp/90.2.271} {\bibfield  {journal}
  {\bibinfo  {journal} {Progr. Theor. Exp. Phys.}\ }\textbf {\bibinfo {volume}
  {90}},\ \bibinfo {pages} {271} (\bibinfo {year} {1993})}\BibitemShut
  {NoStop}%
\bibitem [{\citenamefont {{L. de Broglie}}(1924)}]{deBroglie}%
  \BibitemOpen
  \bibfield  {author} {\bibinfo {author} {\bibnamefont {{L. de Broglie}}},\
  }\emph {\bibinfo {title} {Recherches sur la th{\'e}orie des Quanta}},\
  \href@noop {} {Ph.D. thesis},\ \bibinfo  {school} {University of Paris}
  (\bibinfo {year} {1924})\BibitemShut {NoStop}%
\bibitem [{\citenamefont {Lan}\ \emph {et~al.}(2013)\citenamefont {Lan} \emph
  {et~al.}}]{LanScience}%
  \BibitemOpen
  \bibfield  {author} {\bibinfo {author} {\bibfnamefont {S.-Y.}\ \bibnamefont
  {Lan}} \emph {et~al.},\ }\href@noop {} {\bibfield  {journal} {\bibinfo
  {journal} {Science}\ }\textbf {\bibinfo {volume} {339}},\ \bibinfo {pages}
  {554} (\bibinfo {year} {2013})}\BibitemShut {NoStop}%
\bibitem [{\citenamefont {{Watanabe}}\ and\ \citenamefont
  {{Oshikawa}}(2015)}]{Watanabe}%
  \BibitemOpen
  \bibfield  {author} {\bibinfo {author} {\bibfnamefont {H.}~\bibnamefont
  {{Watanabe}}}\ and\ \bibinfo {author} {\bibfnamefont {M.}~\bibnamefont
  {{Oshikawa}}},\ }\href@noop {} {\bibfield  {journal} {\bibinfo  {journal}
  {Phys. Rev. Lett.}\ }\textbf {\bibinfo {volume} {{\bf 114}}},\ \bibinfo
  {pages} {251603} (\bibinfo {year} {2015})}\BibitemShut {NoStop}%
\bibitem [{\citenamefont {Brody}(2014)}]{BOQM}%
  \BibitemOpen
  \bibfield  {author} {\bibinfo {author} {\bibfnamefont {D.~C.}\ \bibnamefont
  {Brody}},\ }\href {http://stacks.iop.org/1751-8121/47/i=3/a=035305}
  {\bibfield  {journal} {\bibinfo  {journal} {J. Phys. A: Math. Theor.}\
  }\textbf {\bibinfo {volume} {47}},\ \bibinfo {pages} {035305} (\bibinfo
  {year} {2014})}\BibitemShut {NoStop}%
\bibitem [{\citenamefont {Weinberg}(2005)}]{Weinberg}%
  \BibitemOpen
  \bibfield  {author} {\bibinfo {author} {\bibfnamefont {S.}~\bibnamefont
  {Weinberg}},\ }\href
  {https://www.amazon.com/Quantum-Theory-Fields-Foundations/dp/0521670535?SubscriptionId=0JYN1NVW651KCA56C102&tag=techkie-20&linkCode=xm2&camp=2025&creative=165953&creativeASIN=0521670535}
  {\emph {\bibinfo {title} {The Quantum Theory of Fields, Volume 1:
  Foundations}}}\ (\bibinfo  {publisher} {Cambridge University Press},\
  \bibinfo {year} {2005})\BibitemShut {NoStop}%
\bibitem [{\citenamefont {Robertson}(1929)}]{RobertsonUR}%
  \BibitemOpen
  \bibfield  {author} {\bibinfo {author} {\bibfnamefont {H.~P.}\ \bibnamefont
  {Robertson}},\ }\href {\doibase 10.1103/PhysRev.34.163} {\bibfield  {journal}
  {\bibinfo  {journal} {Phys. Rev.}\ }\textbf {\bibinfo {volume} {34}},\
  \bibinfo {pages} {163} (\bibinfo {year} {1929})}\BibitemShut {NoStop}%
\bibitem [{\citenamefont {Dou}\ and\ \citenamefont {Du}(2013)}]{DouDu}%
  \BibitemOpen
  \bibfield  {author} {\bibinfo {author} {\bibfnamefont {Y.-N.}\ \bibnamefont
  {Dou}}\ and\ \bibinfo {author} {\bibfnamefont {H.-K.}\ \bibnamefont {Du}},\
  }\href {\doibase 10.1063/1.4825114} {\bibfield  {journal} {\bibinfo
  {journal} {J. Math. Phys.}\ }\textbf {\bibinfo {volume} {54}},\ \bibinfo
  {pages} {103508} (\bibinfo {year} {2013})}\BibitemShut {NoStop}%
\bibitem [{We note that $\mathcal{PT}$-symmetric operators are not necessarily
  non-Hermitian: For instance()}]{note2}%
  \BibitemOpen
  We note that $\mathcal{PT}$-symmetric operators are not necessarily
  non-Hermitian: For instance,\ \href@noop {} {}\bibinfo {note} {$\hat p^2$ is
  a $\mathcal{PT}$-symmetric Hermitian operator.}\BibitemShut {Stop}%
\bibitem [{\citenamefont {Luminet}\ \emph {et~al.}(2003)\citenamefont
  {Luminet}, \citenamefont {Weeks}, \citenamefont {Riazuelo}, \citenamefont
  {Lehoucq},\ and\ \citenamefont {Uzan}}]{Luminet2003}%
  \BibitemOpen
  \bibfield  {author} {\bibinfo {author} {\bibfnamefont {J.-P.}\ \bibnamefont
  {Luminet}}, \bibinfo {author} {\bibfnamefont {J.~R.}\ \bibnamefont {Weeks}},
  \bibinfo {author} {\bibfnamefont {A.}~\bibnamefont {Riazuelo}}, \bibinfo
  {author} {\bibfnamefont {R.}~\bibnamefont {Lehoucq}}, \ and\ \bibinfo
  {author} {\bibfnamefont {J.-P.}\ \bibnamefont {Uzan}},\ }\href@noop {}
  {\bibfield  {journal} {\bibinfo  {journal} {Nature}\ }\textbf {\bibinfo
  {volume} {425}},\ \bibinfo {pages} {593} (\bibinfo {year}
  {2003})}\BibitemShut {NoStop}%
\bibitem [{\citenamefont {Tanda}\ \emph {et~al.}(2002)\citenamefont {Tanda},
  \citenamefont {Tsuneta}, \citenamefont {Okajima}, \citenamefont {Inagaki},
  \citenamefont {Yamaya},\ and\ \citenamefont {Hatakenaka}}]{ring}%
  \BibitemOpen
  \bibfield  {author} {\bibinfo {author} {\bibfnamefont {S.}~\bibnamefont
  {Tanda}}, \bibinfo {author} {\bibfnamefont {T.}~\bibnamefont {Tsuneta}},
  \bibinfo {author} {\bibfnamefont {Y.}~\bibnamefont {Okajima}}, \bibinfo
  {author} {\bibfnamefont {K.}~\bibnamefont {Inagaki}}, \bibinfo {author}
  {\bibfnamefont {K.}~\bibnamefont {Yamaya}}, \ and\ \bibinfo {author}
  {\bibfnamefont {N.}~\bibnamefont {Hatakenaka}},\ }\href
  {http://dx.doi.org/10.1038/417397a} {\bibfield  {journal} {\bibinfo
  {journal} {Nature}\ }\textbf {\bibinfo {volume} {417}},\ \bibinfo {pages}
  {397 EP } (\bibinfo {year} {2002})}\BibitemShut {NoStop}%
\bibitem [{\citenamefont {Povie}\ \emph {et~al.}(2017)\citenamefont {Povie},
  \citenamefont {Segawa}, \citenamefont {Nishihara}, \citenamefont {Miyauchi},\
  and\ \citenamefont {Itami}}]{Povie172}%
  \BibitemOpen
  \bibfield  {author} {\bibinfo {author} {\bibfnamefont {G.}~\bibnamefont
  {Povie}}, \bibinfo {author} {\bibfnamefont {Y.}~\bibnamefont {Segawa}},
  \bibinfo {author} {\bibfnamefont {T.}~\bibnamefont {Nishihara}}, \bibinfo
  {author} {\bibfnamefont {Y.}~\bibnamefont {Miyauchi}}, \ and\ \bibinfo
  {author} {\bibfnamefont {K.}~\bibnamefont {Itami}},\ }\href {\doibase
  10.1126/science.aam8158} {\bibfield  {journal} {\bibinfo  {journal}
  {Science}\ }\textbf {\bibinfo {volume} {356}},\ \bibinfo {pages} {172}
  (\bibinfo {year} {2017})}\BibitemShut {NoStop}%
\bibitem [{\citenamefont {Li}\ \emph {et~al.}(2017)\citenamefont {Li},
  \citenamefont {Urban}, \citenamefont {Noel}, \citenamefont {Chuang},
  \citenamefont {Xia}, \citenamefont {Ransford}, \citenamefont {Hemmerling},
  \citenamefont {Wang}, \citenamefont {Li}, \citenamefont {H\"affner},\ and\
  \citenamefont {Zhang}}]{Haffner}%
  \BibitemOpen
  \bibfield  {author} {\bibinfo {author} {\bibfnamefont {H.-K.}\ \bibnamefont
  {Li}}, \bibinfo {author} {\bibfnamefont {E.}~\bibnamefont {Urban}}, \bibinfo
  {author} {\bibfnamefont {C.}~\bibnamefont {Noel}}, \bibinfo {author}
  {\bibfnamefont {A.}~\bibnamefont {Chuang}}, \bibinfo {author} {\bibfnamefont
  {Y.}~\bibnamefont {Xia}}, \bibinfo {author} {\bibfnamefont {A.}~\bibnamefont
  {Ransford}}, \bibinfo {author} {\bibfnamefont {B.}~\bibnamefont
  {Hemmerling}}, \bibinfo {author} {\bibfnamefont {Y.}~\bibnamefont {Wang}},
  \bibinfo {author} {\bibfnamefont {T.}~\bibnamefont {Li}}, \bibinfo {author}
  {\bibfnamefont {H.}~\bibnamefont {H\"affner}}, \ and\ \bibinfo {author}
  {\bibfnamefont {X.}~\bibnamefont {Zhang}},\ }\href@noop {} {\bibfield
  {journal} {\bibinfo  {journal} {Phys. Rev. Lett.}\ }\textbf {\bibinfo
  {volume} {118}},\ \bibinfo {pages} {053001} (\bibinfo {year}
  {2017})}\BibitemShut {NoStop}%
\bibitem [{\citenamefont {Ustinov}\ \emph {et~al.}(1992)\citenamefont
  {Ustinov}, \citenamefont {Doderer}, \citenamefont {Huebener}, \citenamefont
  {Pedersen}, \citenamefont {Mayer},\ and\ \citenamefont
  {Oboznov}}]{AnnularSoliton}%
  \BibitemOpen
  \bibfield  {author} {\bibinfo {author} {\bibfnamefont {A.~V.}\ \bibnamefont
  {Ustinov}}, \bibinfo {author} {\bibfnamefont {T.}~\bibnamefont {Doderer}},
  \bibinfo {author} {\bibfnamefont {R.~P.}\ \bibnamefont {Huebener}}, \bibinfo
  {author} {\bibfnamefont {N.~F.}\ \bibnamefont {Pedersen}}, \bibinfo {author}
  {\bibfnamefont {B.}~\bibnamefont {Mayer}}, \ and\ \bibinfo {author}
  {\bibfnamefont {V.~A.}\ \bibnamefont {Oboznov}},\ }\href@noop {} {\bibfield
  {journal} {\bibinfo  {journal} {Phys. Rev. Lett.}\ }\textbf {\bibinfo
  {volume} {69}},\ \bibinfo {pages} {1815} (\bibinfo {year}
  {1992})}\BibitemShut {NoStop}%
\bibitem [{\citenamefont {Erhart}\ \emph {et~al.}(2012)\citenamefont {Erhart},
  \citenamefont {Sponar}, \citenamefont {Sulyok}, \citenamefont {Badurek},
  \citenamefont {Ozawa},\ and\ \citenamefont {Hasegawa}}]{Hasegawa}%
  \BibitemOpen
  \bibfield  {author} {\bibinfo {author} {\bibfnamefont {J.}~\bibnamefont
  {Erhart}}, \bibinfo {author} {\bibfnamefont {S.}~\bibnamefont {Sponar}},
  \bibinfo {author} {\bibfnamefont {G.}~\bibnamefont {Sulyok}}, \bibinfo
  {author} {\bibfnamefont {G.}~\bibnamefont {Badurek}}, \bibinfo {author}
  {\bibfnamefont {M.}~\bibnamefont {Ozawa}}, \ and\ \bibinfo {author}
  {\bibfnamefont {Y.}~\bibnamefont {Hasegawa}},\ }\href@noop {} {\bibfield
  {journal} {\bibinfo  {journal} {Nature Physics}\ }\textbf {\bibinfo {volume}
  {8}},\ \bibinfo {pages} {185} (\bibinfo {year} {2012})}\BibitemShut {NoStop}%
\end{thebibliography}%
\end{document}